\documentclass[10pt]{iopart}
\usepackage{epsfig}
\begin{document}

\title[Large-scale weighted communication network]{Analysis of a large-scale weighted network of one-to-one human communication}

\author{Jukka-Pekka Onnela$^{1,2}$}
\author{Jari Saram\"aki$^1$}
\author{J\"orkki Hyv\"onen$^1$}
\author{G\'{a}bor Szab\'{o}$^{3,4}$}
\author{M. Argollo de Menezes$^3$}
\author{Kimmo Kaski$^1$}
\author{Albert-L\'{a}szl\'{o} Barab\'{a}si$^{3,4}$}
\author{J\'anos Kert\'esz$^{1,5}$}

\vspace{1cm}

\address{$^1$ Laboratory of Computational Engineering, Helsinki University of Technology, Finland}
\address{$^2$Clarendon Laboratory, Physics Department, Oxford University, Oxford, U.K}
\address{$^3$Department of Physics and Center for Complex Networks Research, University of Notre Dame, IN, USA}
\address{$^4$Center for Cancer Systems Biology, Dana Farber Cancer Institute, Harvard University, Boston, MA, USA}
\address{$^5$Department of Theoretical Physics, Budapest University of Technology and Economics, Budapest, Hungary}

\begin{abstract}
We construct a connected network of 3.9 million nodes from mobile phone call records, which can be regarded as a proxy for the underlying human communication network at the societal level. We assign two weights on each edge to reflect the strength of social interaction, which are the aggregate call duration and the cumulative number of calls placed between the individuals over a period of 18 weeks. We present a detailed analysis of this weighted network by examining its degree, strength, and weight distributions, as well as its topological assortativity and weighted assortativity, clustering and weighted clustering, together with correlations between these quantities. We give an account of motif intensity and coherence distributions and compare them to a randomized reference system. We also use the concept of link overlap to measure the number of common neighbors any two adjacent nodes have, which serves as a useful local measure for identifying the interconnectedness of  communities. We report a positive correlation between the overlap and weight of a link, thus providing strong quantitative evidence for the weak ties hypothesis, a central concept in social network analysis. The percolation properties of the network are found to depend on the type and order of removed links, and they can help understand how the local structure of the network manifests itself at the global level. We hope that our results will contribute to modeling weighted large-scale social networks, and believe that the systematic approach followed here can be adopted to study other weighted networks.
\end{abstract}

\section{Introduction \& Data} \label{sec:intro}

Social networks have been a subject of intensive study since the 1930's. In this framework social life consists of the flow and exchange of norms, values, ideas, and other social and cultural resources \cite{white:1976}, and social action of individuals is affected by the structure of the underlying network \cite{granovetter:1992}. The structure of social networks is important then not only from the perspective of the individual, but also from that of the society as a whole. However, uncovering the structure of social networks has been constrained by the practical difficulty of mapping out interactions among a large number of individuals. Social scientists have ordinarily based their studies on questionnaire data, typically reaching the order of $N \approx 10^2$ individuals \cite{wasserman:1994}. Although the spectrum of social interactions that may be probed in this approach is wide, the strength of an interaction is often based on recollection and, consequently, is highly subjective. However, in the late the 1990's a change of paradigm took place \cite{watts:1998,albert:1999}. Physicists became interested in large scale social networks, utilizing electronic databases from emails \cite{ebel:2002,eckmann:2004,dodds:2003} to phone records \cite{aiello:2000}, offering unprecedented opportunities to uncover and explore large-scale social networks \cite{watts:2007}. In this scheme the order of $N \approx 10^6$ individuals may be handled and, although the range of social interactions is narrower, in some cases their strengths may be objectively quantifiable. While both approaches have their merits, studying large-scale networks has potential to shed light on how individual microscopic interactions translate into macroscopic social systems. In addition to this being one of the key questions as posed by social scientists in the field, it is also the one to which statistical physics in general, and the science of complex networks in particular, can make a contribution.

In this paper we present a detailed analysis of a network constructed from a data set consisting of the mobile phone call records of over seven million individuals over a period of 18 weeks (126 days), covering approximately 20\% of the population of the country. For the purpose of retaining customer anonymity, each subscription was identified by a surrogate key, guaranteeing that the privacy of customers was respected. We kept only voice calls, filtering out all other services, such as voice mail, data calls, text messages, chat, and operator calls. We filtered out calls involving other operators, incoming or outgoing, keeping only those transactions in which the calling and receiving subscription is governed by the same operator. This filtering was needed to eliminate the bias between this operator and other operators as we have a full access to the call records of this operator, but only partial access to the call records of other operators. We constructed two different networks from the data. In the first scheme we connected two users with an undirected link if there had been at least one phone call between them, i.e., $i$ called $j$ \emph{or} $j$ called $i$, resulting in a \emph{non-mutual network} consisting of $N=7.2 \times 10^6$ nodes and $L=22.6 \times 10^6$ links. However, many of these calls are one-way, most of which correspond to single events, suggesting that they typically reach individuals that the caller might not know personally. To eliminate them, in the second scheme we connected two users with an undirected link if there had been at least one reciprocated pair of phone calls between them, i.e., $i$ called $j$ \emph{and} $j$ called $i$, resulting in a \emph{mutual network} with $N=4.6 \times 10^6$ nodes and $L=7.0 \times 10^6$ links.

The resulting mobile call graph (MCG) naturally captures only a subset of the underlying social network, which consists of all forms of social interactions, including face-to-face interactions, email and landline communication etc. However, research on media multiplexity suggests that the use of one medium for communication between two people implies communication via other means as well \cite{haythornthwaite:2005}. Furthermore, in the absence of directory listings, the mobile phone data is skewed towards trusted interactions, i.e., people tend to share their mobile numbers only with individuals they trust. Therefore, the MCG can be used as a proxy for the underlying social network. 

We can quantify the weight of the link $(i,j)$ by the aggregated time $i$ and $j$ spent talking to each other as well as by the total number of calls made between $i$ and $j$ over the studied period. These weights are denoted by $w_{ij}^D$ (total duration of calls) and $w_{ij}^N$ (total number of calls), respectively, where the former is measured in seconds ($s$) and the latter is a dimensionless quantity.

This paper is devoted to the study of these weighted, large-scale, one-to-one social interaction networks, with emphasis on the mutual over the non-mutual network. We adopt a "cookbook approach" by carrying out a systematic analysis of basic and more advanced network characteristics, and hope that others working on weighted networks will benefit from our "recipes". We study some of the basic network  characteristics in Section \ref{sec:basic} and focus on weighted network characteristics in Section \ref{sec:adv}. We explore the coupling between link weight and the surrounding local network topology in Section \ref{sec:link}. We have dedicated Section \ref{sec:perc} to the study of percolation properties of the network and, finally, discuss our findings in Section \ref{sec:disc}.

\section{Basic network characteristics}
\label{sec:basic}

\begin{figure}
\begin{center}
\includegraphics[width=10.0cm]{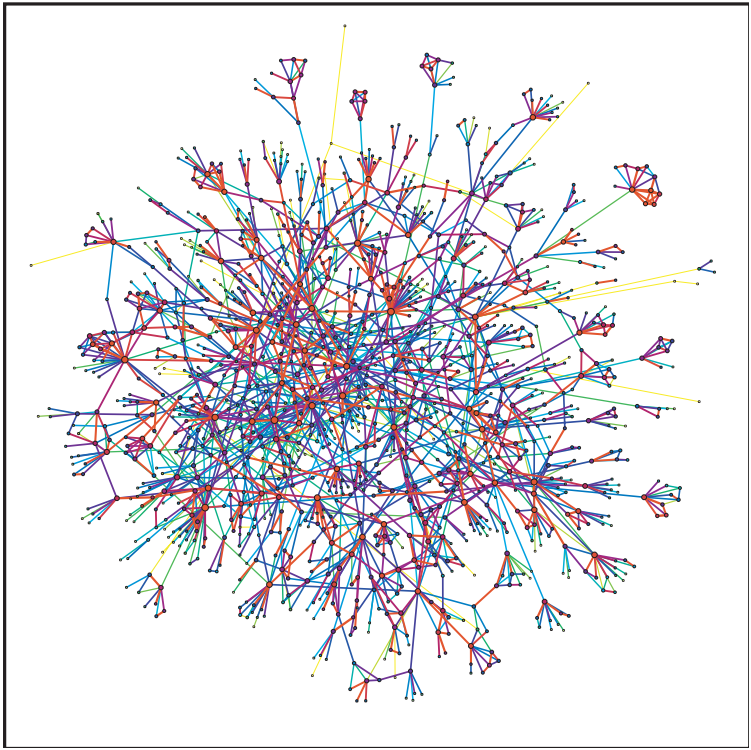}
\caption{A small sample of the network with link weights $w_{ij}^D$ color coded from yellow (weak link) to red (strong link). }
\label{fig:sample}
\end{center}
\end{figure}

We start inspecting the network by showing a small sample of it in Fig.~\ref{fig:sample}. The sample has been extracted from the mutual network by picking a node (source node) at random and including all nodes in the sample that are within a (topological) distance of $\ell = 5$ from the source node. This method of sampling is sometimes called snowball sampling \cite{lee:2006}. The color of links corresponds to the strength of each tie in terms of $w_{ij}^D$. It appears from this figure that the network consists of small local clusters, and the majority of the strong ties (colored in red) seem to be localized within these clusters. In some cases nodes connected by a strong link have many common neighbors, but there are also strongly connected nodes with few or no common neighbors. 

These two apparently contradictory trends arise as a result of being forced to examine a sample of the network as opposed to the entire network. To understand the limits of visual inspection, it is important to realize that since the network is a high dimensional object, a majority of the nodes will be on the outskirts of the sample. A consequence of this is that for most of these nodes we only have partial visibility into their neighborhood. Consequently, one can see the full neighborhood for only a small minority of nodes in the sample.

\begin{figure}
\begin{center}
\includegraphics[width=8.0cm]{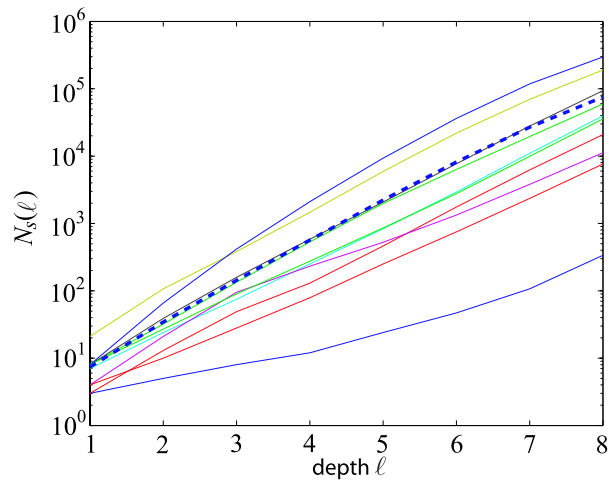}
\caption{Number of nodes in the sample $N_s(\ell)$, obtained by snowball sampling, as a function of extraction distance $\ell$ for several choices of the source node (solid lines) and their average (dashed line). }
\label{fig:extraction}
\end{center}
\end{figure}

Let us elaborate on network sampling. We show in Fig.~\ref{fig:extraction} the number of nodes in the sample $N_s(\ell)$, obtained using snowball sampling, as a function of extraction distance $\ell$ for several choices of the source node (solid lines) and their average (dashed line). Here $N_s(\ell)$ is defined as the number of nodes within a distance $\ell$ from the given source node. For a fixed value of $x$, we call nodes for which $\ell < x$ \emph{bulk nodes} and those with $\ell = x$ \emph{surface nodes} of the sample. The number of surface nodes clearly outweighs the number of bulk nodes. This is to be expected since the network behaves like a high dimensional hypersphere, the volume of which is negligible to its surface area. To a good approximation we can write $N_s = A e^{B \ell}$, where $A$ and $B$ are fitting parameters. In general, the number of of surface nodes to the number of bulk nodes is $[N_s(\ell) - N_s(\ell - 1)] / N_s(\ell) = 1 - e^{-B}$. Thus, a large majority of nodes are surface nodes.

\begin{figure}
\begin{center}
\includegraphics[width=10.0cm]{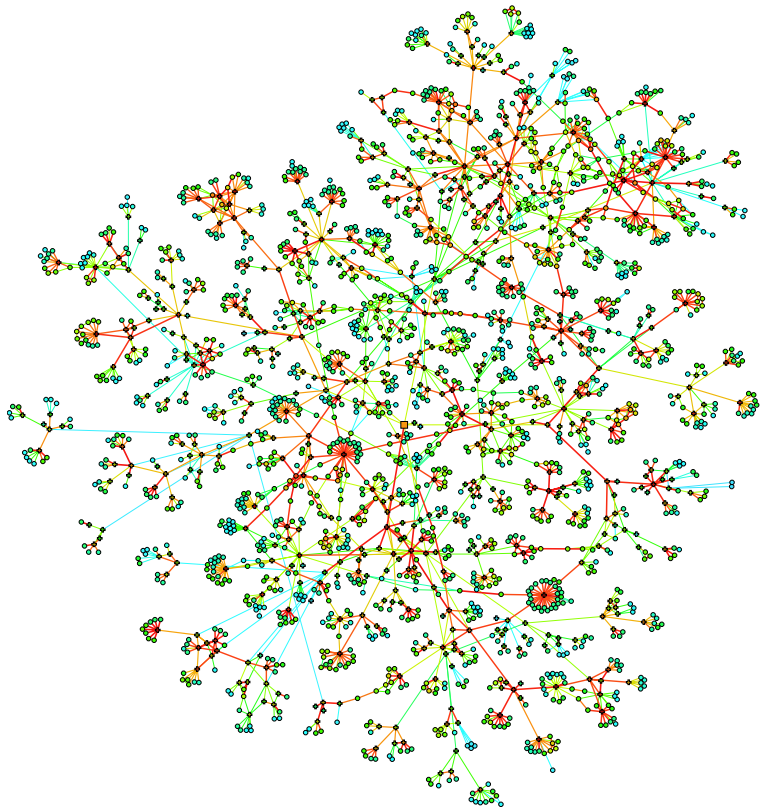}
\caption{A sample of the network, showing the source (square at the center) node from which sampling was started, the bulk nodes($+$), and surface nodes ($\circ$). For surface nodes, which clearly are in the majority, only some of their nearest neighbors are visible in the sample, while the rest are outside the sample.}
\label{fig:surfacenodes}
\end{center}
\end{figure}

This is clear from another network sample in Fig.~\ref{fig:surfacenodes}, in which bulk nodes and surface nodes are drawn with different markers. It is only for bulk nodes to which we have full visibility of their neighborhood and, consequently, may make unbiased judgments about the structure of their neighborhoods. Since these nodes are clearly in the minority, it is clear that visual inspection of network samples has limited utility. 

\begin{figure}
\begin{center}
\includegraphics[width=0.45\linewidth]{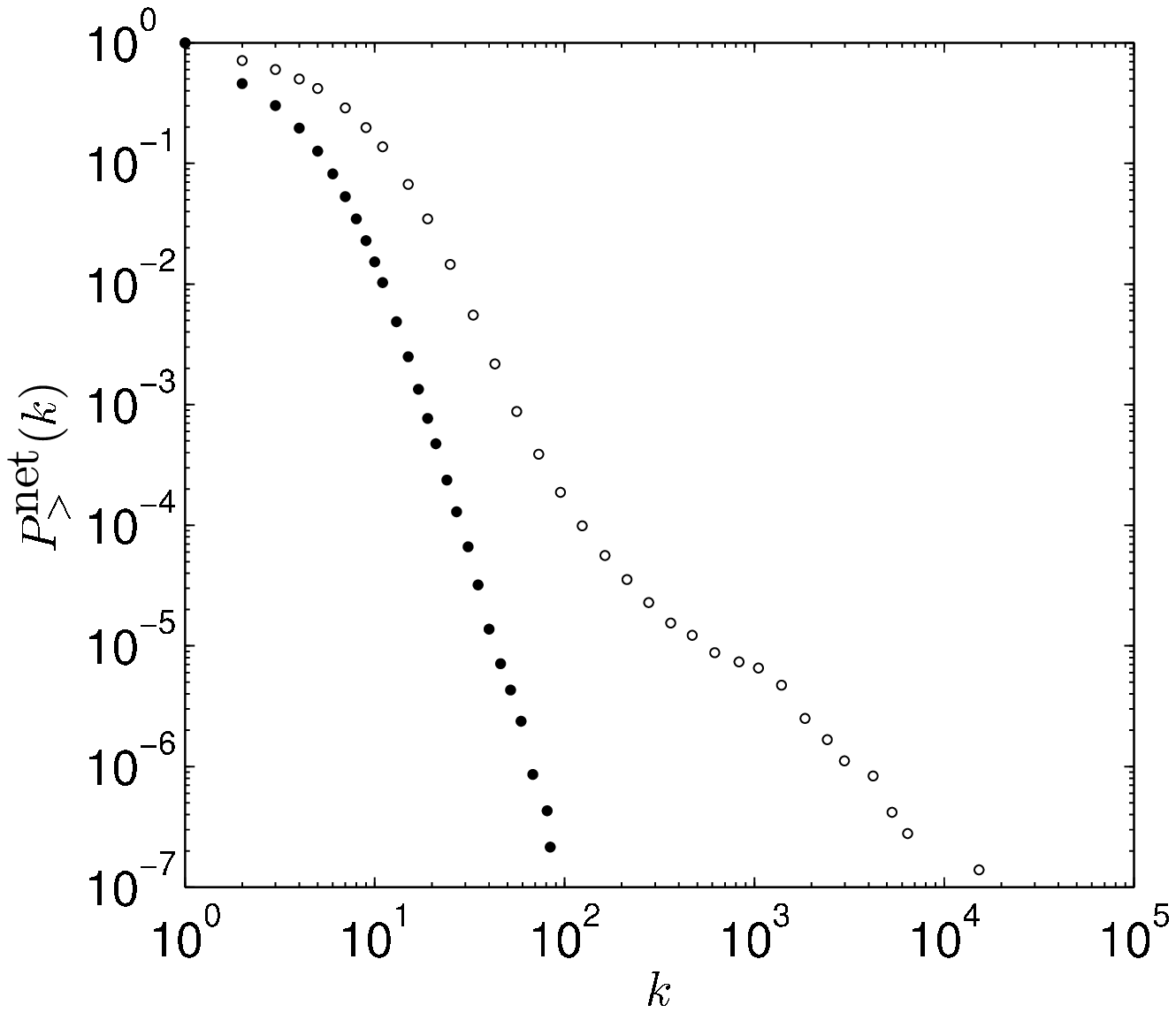}
\includegraphics[width=0.50\linewidth]{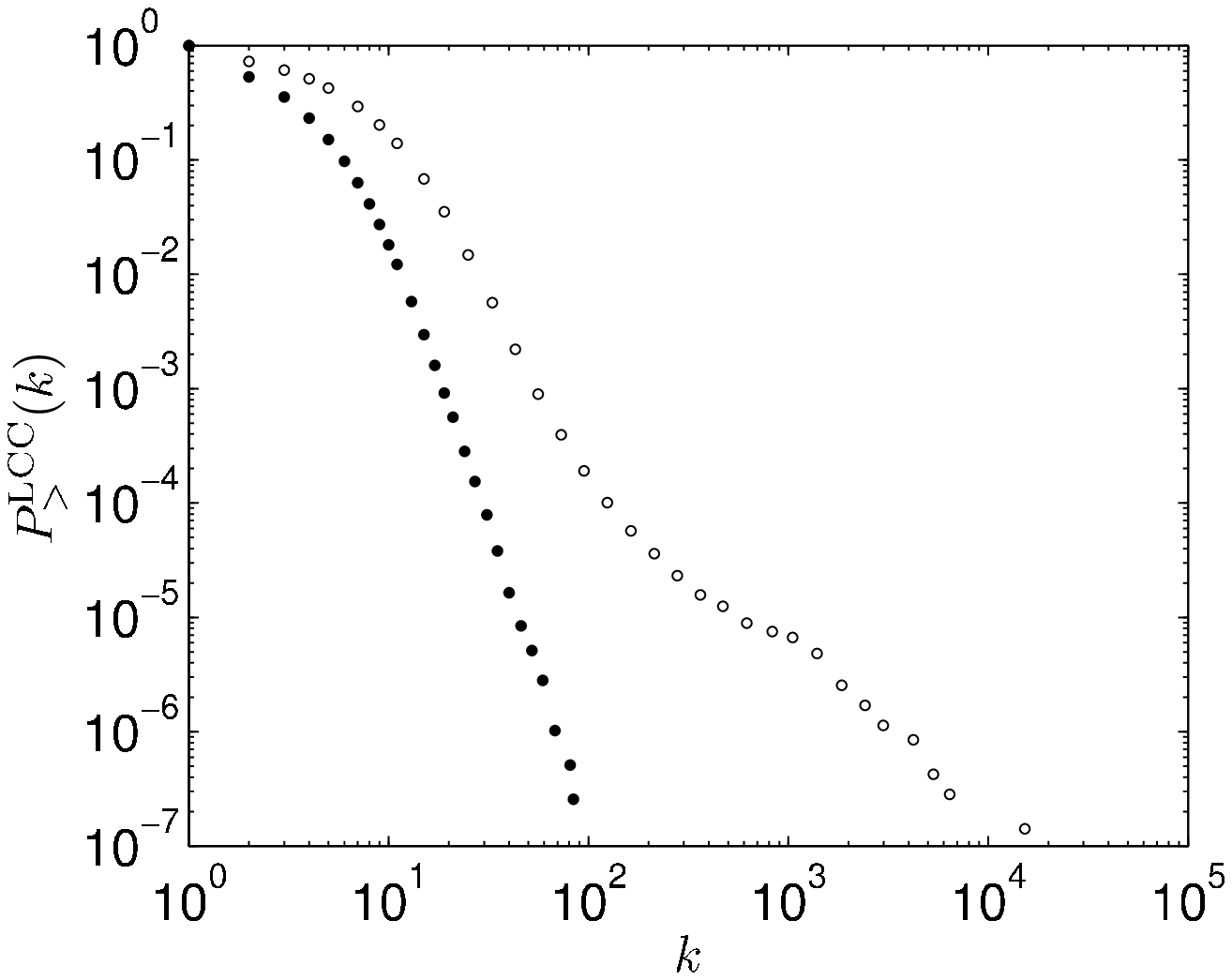}
\caption{Cumulative degree distribution $P_{>}^{\textrm{net}}(k)$ for the mutual ($\bullet$) and non-mutual ($\circ$) networks (left) and for their respective largest connected components $P_{>}^{\textrm{LCC}}(k)$ (right). The mutual network is a subgraph of the non-mutual one, and 84.1\% of the nodes in the mutual network belong to a single connected component (LCC), for which the average degree $\langle k \rangle \approx 3.0$.}
\label{fig:degree}
\end{center}
\end{figure}

A basic network characteristic, the degree distribution, is shown in Fig.~\ref{fig:degree}. To avoid the need of binning, we study the cumulative degree distribution $P_{>}(k)$, defined as $P_{>}(k) = \int_{k}^{\infty} p(x) \, dx$, where $p(x)$ the degree probability density function. We denote the distribution for whole mutual and non-mutual networks $P_{>}^{\textrm{net}}(k)$, and that of their respective largest connected components (LCC) by $P_{>}^{\textrm{LCC}}(k)$. Note that the mutual network is a subgraph of the non-mutual one, and the LCC is a subgraph of the whole network. In the case of the mutual network 84\% of the nodes belong to the LCC. In this case little is left outside the LCC, partly explaining why distributions are almost identical for the whole network and the LCC. 

In general, the degree distributions are skewed with a fat tail, indicating that while most users communicate with only a few individuals, a small minority talks with dozens. The noticeable difference between the degree distributions for the mutual and non-mutual network is the fatter tail of the non-mutual network. In particular, the non-mutual network has a fatter tail, so that while the most connected node in the LCC of the mutual network has $k_{\max} = 144$, in the LCC of the non-mutual network $k_{\max} = 34625$. Clearly, the latter cannot correspond to a single individual. However, it appears plausible that the mutual network is dominated by trusted interactions, i.e., people tend to share their mobile numbers only with individuals they trust. We also point out that $k_{\max} = 144$ in the mutual network is very close to the approximate number of $k=150$ put forward by Dunbar as a limit on connectivity resulting from the size of neocortex in the cerebral cortex in primates \cite{dunbar:1992}. From now on, unless otherwise mentioned, we shall focus exclusively on the LCC of the mutual network.

The tail of the degree distribution $P(k)$ for the LCC of the mutual network is approximated well by a power law of the form $P(k) = a(k+k_0)^{-\gamma}$ with $k_0=10.9$ and $\gamma=8.4$. Note that the value of the exponent is significantly higher than the value observed for landlines ($\gamma = 2.1$ for the in-degree distribution \cite{dorogovtesev:2003}). For such a rapidly decaying degree distribution the hubs are few, and therefore many properties of traditional scale-free networks, from anomalous diffusion \cite{pastor-satorras:2001a} to error tolerance \cite{cohen:2000}, are absent.

\begin{figure}
\begin{center}
\includegraphics[width=0.50\linewidth]{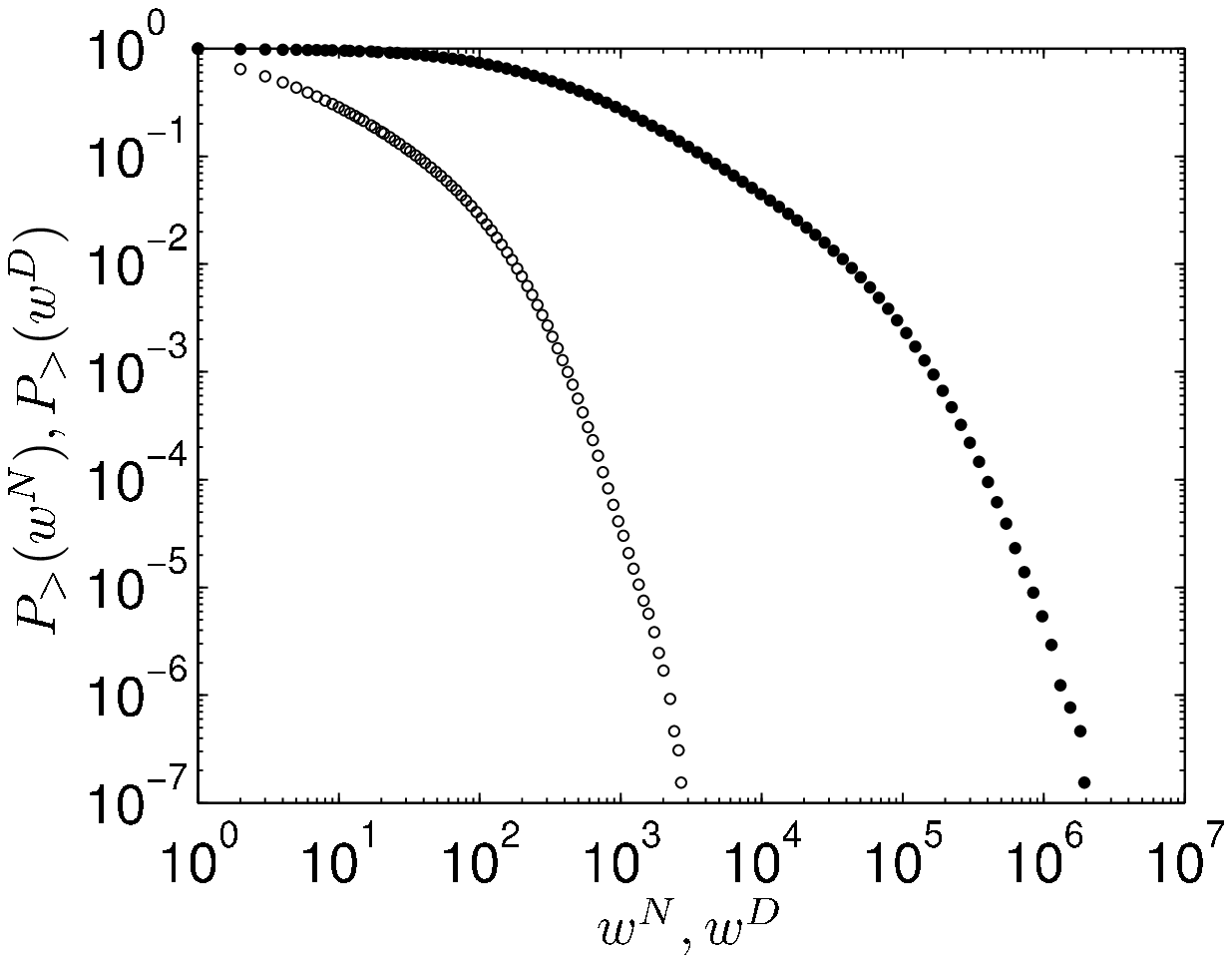}
\includegraphics[width=0.48\linewidth]{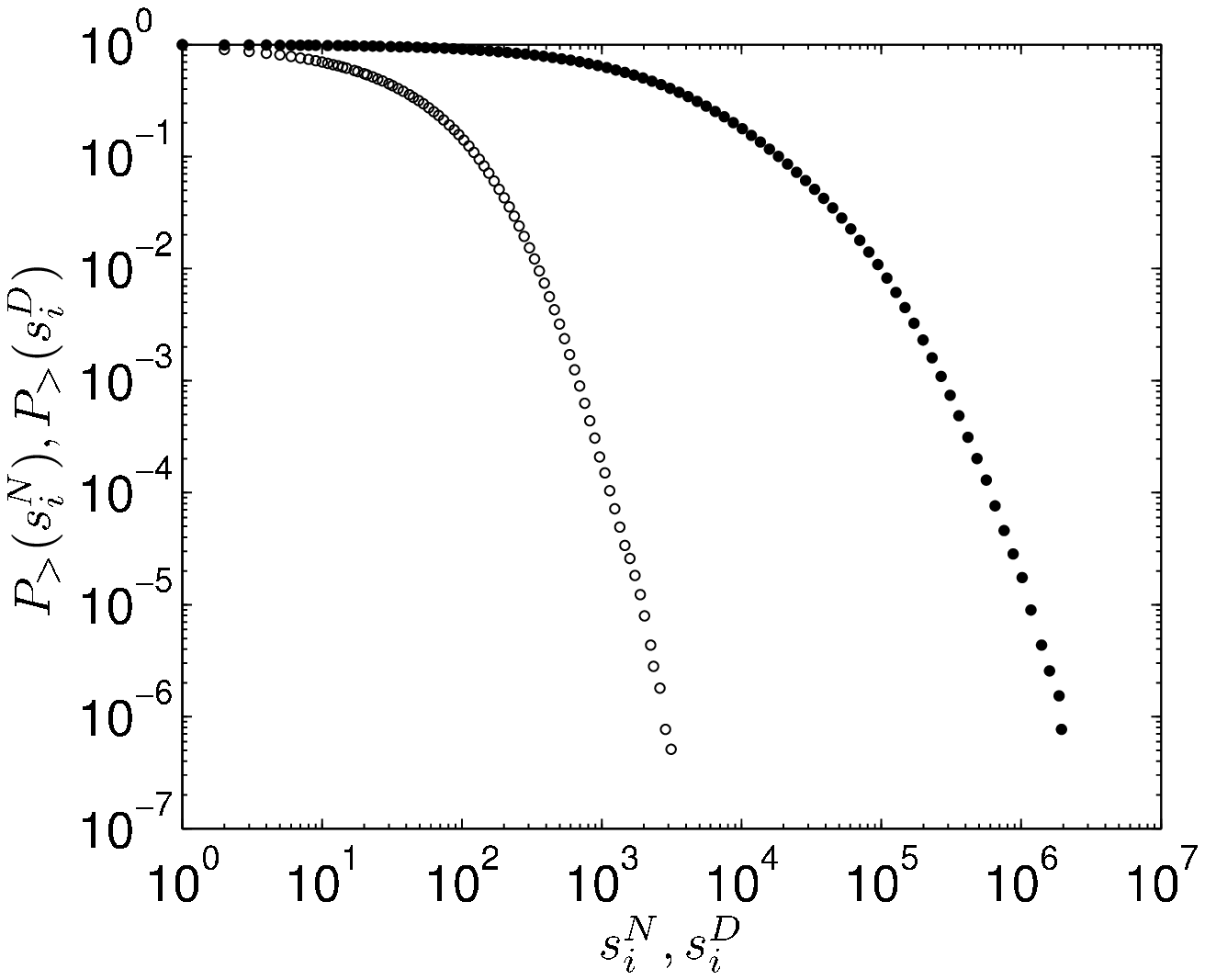}
\caption{Cumulative link weight distributions (left) and cumulative node strength distributions (right) in the LCC of the mutual network. Link weights and node strengths are measured in terms of the absolute number of calls made during the studied period ($\circ$), corresponding to $P_{>}^{N}(w)$ and $P_{>}^{N}(s)$, as well as the aggregated call duration during the period ($\bullet$), given by $P_{>}^{D}(w)$ and $P_{>}^{D}(s)$.}
\label{fig:weightdist}
\end{center}
\end{figure}

\begin{figure}
\begin{center}
\includegraphics[width=0.6\linewidth]{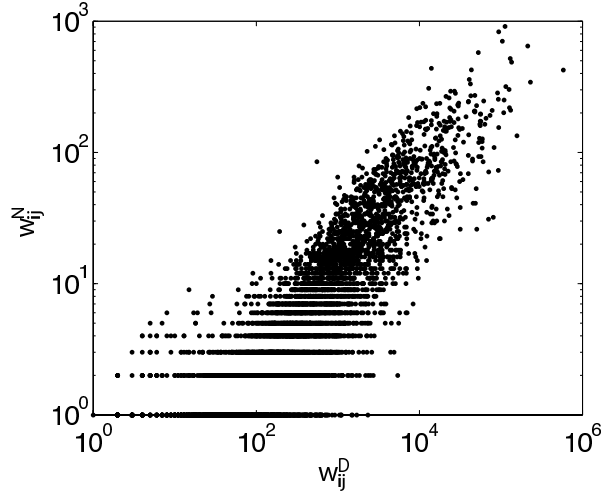}
\caption{Scatter plot of call duration weights $w_{ij}^D$ and number of calls weights $w_{ij}^N$. The two weights are clearly correlated in this random sample of 5000 links, as well as in the LCC of the mutual network, giving rise to Pearson's linear correlation coefficient of $0.70$ in the latter.}
\label{fig:scatter}
\end{center}
\end{figure}

As mentioned in the Introduction, link weights and node strengths are measured in terms of the absolute number of calls made during the studied period. The associated cumulative distributions are $P_{>}^{N}(w)$ and $P_{>}^{N}(s)$ for the number of calls, and $P_{>}^{D}(w)$ and $P_{>}^{D}(s)$ for the  aggregated call duration as shown in Fig.~\ref{fig:weightdist}. Both link weight distributions are broad so that while the majority of ties correspond to a couple of calls and a few minutes of air time, a small fraction of users place numerous calls and spend hours chatting with each other. On average an individual made $\langle s^{N} \rangle \approx 51.1$ calls and spent $\langle s^{D} \rangle \approx 8074 s$ (135 mins) on the phone. Two connected individuals spoke on average $\langle w^{N} \rangle \approx 15.4$ times on the phone spending altogether $\langle w^{D} \rangle \approx 2429 s$ (40 mins) talking to one other. These values are summarized in Table \ref{table:stats}, which also lists some higher moments for the distributions. The two weights $w_{ij}^D$ and $w_{ij}^N$ are strongly correlated as expected, and this is evident in Fig.~\ref{fig:scatter}. In the mutual network Pearson's linear correlation coefficient between $w_{ij}^D$ and $w_{ij}^N$ is $0.70$, implying that variance in $w_{ij}^N$ explains some 50\% of variance in $w{ij}^D$.

The tail of the weight distribution $P(w^D)$ for the LCC of the mutual network is approximated well by an exponentially truncated power-law of the form $P(w) = a(w+w_0)^{-\gamma}\exp{(-w/w_c)}$ with $w_0=280$, $\beta=1.9$, and the cut-off parameter $w_c = 3.4 \times 10^{5}$. The broad tailed nature of these distributions is rather unexpected, given that fat tailed tie strength distributions have been observed mainly in networks characterized by global transport processes, such as the number of passengers carried by the airline transportation network \cite{colizza:2006}, the reaction fluxes in metabolic networks \cite{almaas:2004}, and packet transfer on the Internet \cite{goh:2001}. In all these cases the individual fluxes are determined by the global network topology, in which an important property is "conservation of mass", i.e., local conservation of passengers, molecules, and data packets. Such constraints are not present here and, in addition, social networks are expected to be fairly local in nature, meaning that the nature of the link weight and strength distributions are non-trivial. This raises the interesting question of the extent to which network structure and link weights are correlated in this network and, in general, whether their extent of correlation can be used to categorize networks in different classes. We will address the first part of this question in Section \ref{sec:link}. 

\begin{table}
\begin{center}
\scriptsize
\begin{tabular}[]{|l|l|l|l|l|l|}
\hline
$x$ & mean & std & skewness & kurtosis & max \\
\hline
degree $k_i,\textrm{net, NM}$ & $6.28$             & $16.6$             & $1.39 \times 10^3$ & $2.71 \times 10^6$ & $3.46 \times 10^4$\\
degree $k_i,\textrm{net, M}$  & $3.01$             & $2.41$             & $2.40$             & $17.5$             & $144$\\
degree $k_i,\textrm{LCC, NM}$  & $6.37$             & $16.8$             & $1.38 \times 10^3$ & $2.68 \times 10^6$ & $3.46 \times 10^4$\\
degree $k_i,\textrm{LCC, M}$   & $3.32$             & $2.49$             & $2.28$             & $17.0$             & $144$\\
weight $w_{ij}^N$             & $15.4$             & $37.3$             & $8.54$             & $165$              & $3.61 \times 10^3$\\
weight $w_{ij}^D$             & $2.43 \times 10^3$ & $1.23 \times 10^4$ & $25.1$             & $1.52 \times 10^3$ & $2.39 \times 10^6$\\
strength $s_i^N$                & $51.1$             & $74.8$             & $4.30$             & $44.2$             & $3.64 \times 10^3$\\
strength $s_i^D$                & $8.07 \times 10^3$ & $2.32 \times 10^4$ & $13.5$             & $452$              & $2.48 \times 10^6$\\
\hline
\end{tabular}
\caption{Summary of descriptive network statistics. The following terms are used: whole network (net), largest connected component (LCC), non-mutual network (NM) and mutual network (M). The superscripts $N$ and and $D$ refer to number-of-calls and aggregate-call-duration based weights and strengths, respectively.} 
\label{table:stats}
\end{center}
\end{table}

Social networks are expected to be assortative: People with many friends are connected to others who also have many friends. This gives rise to degree-degree correlations in the network, meaning that the the degrees of two adjacent nodes are not independent. These correlations are completely described by the joint probability distribution $P(k,k')$, giving the probability that a node of degree $k$ is connected to a node of degree $k'$. It is more practical, however, to define the average nearest neighbours degree of a node $v_i$ as $k_{nn,i} = (1/k_i) \sum_{j \in \mathcal{N}(v_i)} k_j$, where $\mathcal{N}(v_i)$ denotes the neighbourhood of $v_i$. By averaging this over all nodes in the network of a given degree $k$, one can calculate the average degree of nearest neighbors with degree $k$ denoted by $\langle k_{nn} | k \rangle$, which corresponds to $\sum_{k'}k'P(k'|k)$ \cite{pastor-satorras:2001}. The network is said to exhibit \emph{assortative mixing} if $\langle k_{nn} | k \rangle$ increases and \emph{disassortative mixing} if it decreases as a function of $k$ \cite{newman:2002}.

\begin{figure}
\begin{center}
\includegraphics[width=0.49\linewidth]{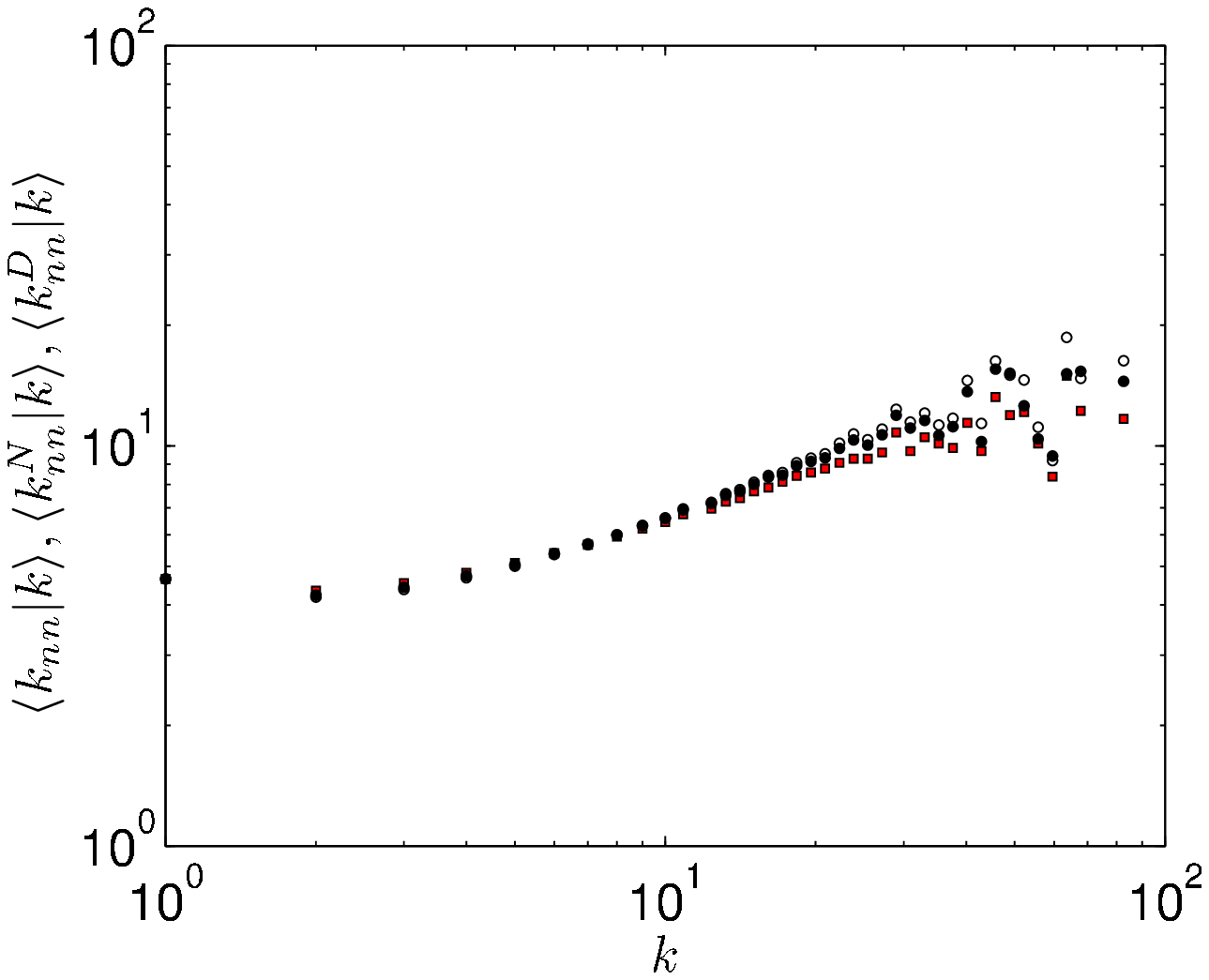}
\includegraphics[width=0.49\linewidth]{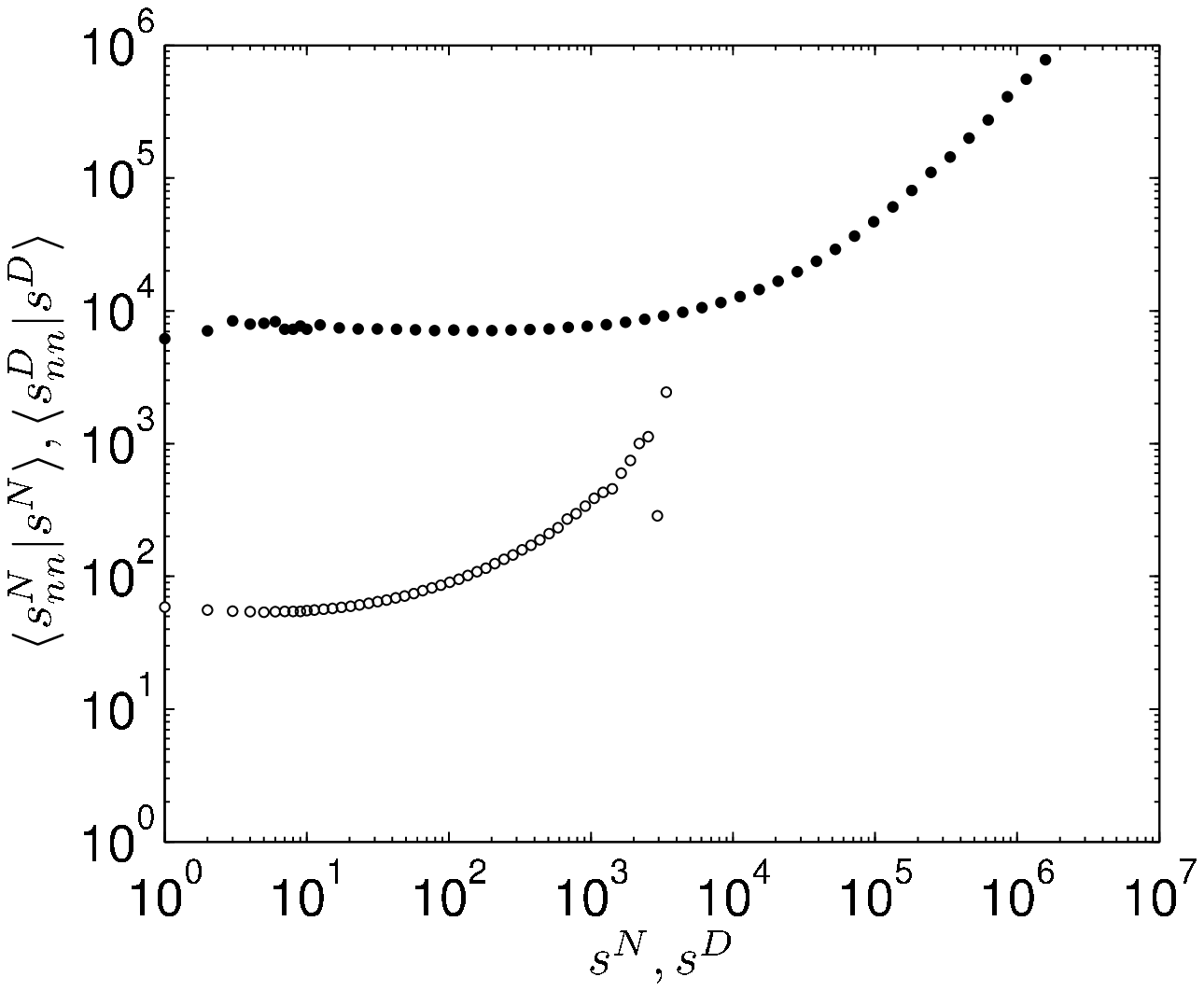}
\caption{Average neighbor degree $\langle k_{nn} | k \rangle $, $\langle k_{nn}^N | k \rangle$, and $\langle k_{nn}^D | k \rangle$ (left) and average neighbor strength $\langle s_{nn}^D | s^D \rangle$ and $\langle s_{nn}^N | s^N \rangle$ (right) in the LCC of the mutual network. The three markers in the plot on the left correspond to unweighted $\langle k_{nn} \rangle$ (black squares), number-of-calls weighted $\langle k_{nn}^N \rangle$ ($\circ$), and call-duration weighted $\langle k_{nn}^D \rangle$ ($\bullet$) averages. The markers on the right correspond to number of calls ($\circ$) and total call duration ($\bullet$).}
\label{fig:neigh}
\end{center}
\end{figure}

We show the average nearest neighbor degree in Fig.~\ref{fig:neigh}. We follow Barrat \emph{et al.} and use the weighted average nearest neighbor degree to characterize degree-degree correlations \cite{barrat:2004}, which are written as $k_{nn,i}^N = (1/s_i^N) \sum_{j \in \mathcal{N}(v_i)} w_{ij}^N k_j$ and $k_{nn,i}^D = (1/s_i^D) \sum_{j \in \mathcal{N}(v_i)} w_{ij}^D k_j$, corresponding to the two weighting schemes. Averaging these over the network gives $\langle k_{nn}|k \rangle$, $\langle k_{nn}^N|k \rangle$ and $\langle k_{nn}^D|k \rangle$, which measure the effective affinity to connect with neighbors of a given degree while taking the magnitude of the interactions into account \cite{barrat:2004}. The three measures behave very similarly in Fig.~\ref{fig:neigh}, and the network is clearly assortative degree-wise such that $\langle k_{nn} | k \rangle \sim k^{\alpha}$ applies with $\alpha \approx 0.4$. 

In addition to degree-degree correlations, which characterize the topology of the network, we can study correlations between node strengths, where node strength is given by $s_i = \sum_{j \in \mathcal{N}(v_i)} w_{ij}$. The average nearest neighbour strengths are given by $s_{nn,i}^N = (1/k_i) \sum_{j \in \mathcal{N}(v_i)} s_j^N$ and $s_{nn,i}^D = (1/k_i) \sum_{j \in \mathcal{N}(v_i)} s_j^D$ which, when averaged over all nodes in the network with strength approximately equal to $s$, gives the average strength of nearest neighbors $\langle s_{nn}^N  | s^N \rangle$ and $\langle s_{nn}^D  | s^D \rangle$. Whereas the degrees of two adjacent nodes are strongly correlated, we find that the strengths of two adjacent nodes in most cases are not. Fig.~\ref{fig:neigh} shows that the $s^D$ dependence of $\langle s_{nn}^D  | s \rangle \sim s^{\alpha^{D}}$ can be divided into two parts, where the independence observed for small $s^D$ crosses over at $s_x  \approx 10^4$ to a linear relationship. This linear region can be understood by studying the the proportion of node strength that is contributed by a single link. It turns out that for very strong links with $w^D > 10^4$, which make up $4.4$ \% of all links, the strength of both adjacent nodes is determined almost entirely by the weight of this single link  such that $s_i \approx w_{ij} \approx s_j$ \cite{onnela:2007}. This explains the linear trend in strength-strength correlations. The plot for $\langle s_{nn}^N  | s^N \rangle$ suggests a qualitatively similar picture, where the linear trend naturally sets in earlier in terms of the absolute value of $s^N$.

\begin{figure}
\begin{center}
\includegraphics[width=0.49\linewidth]{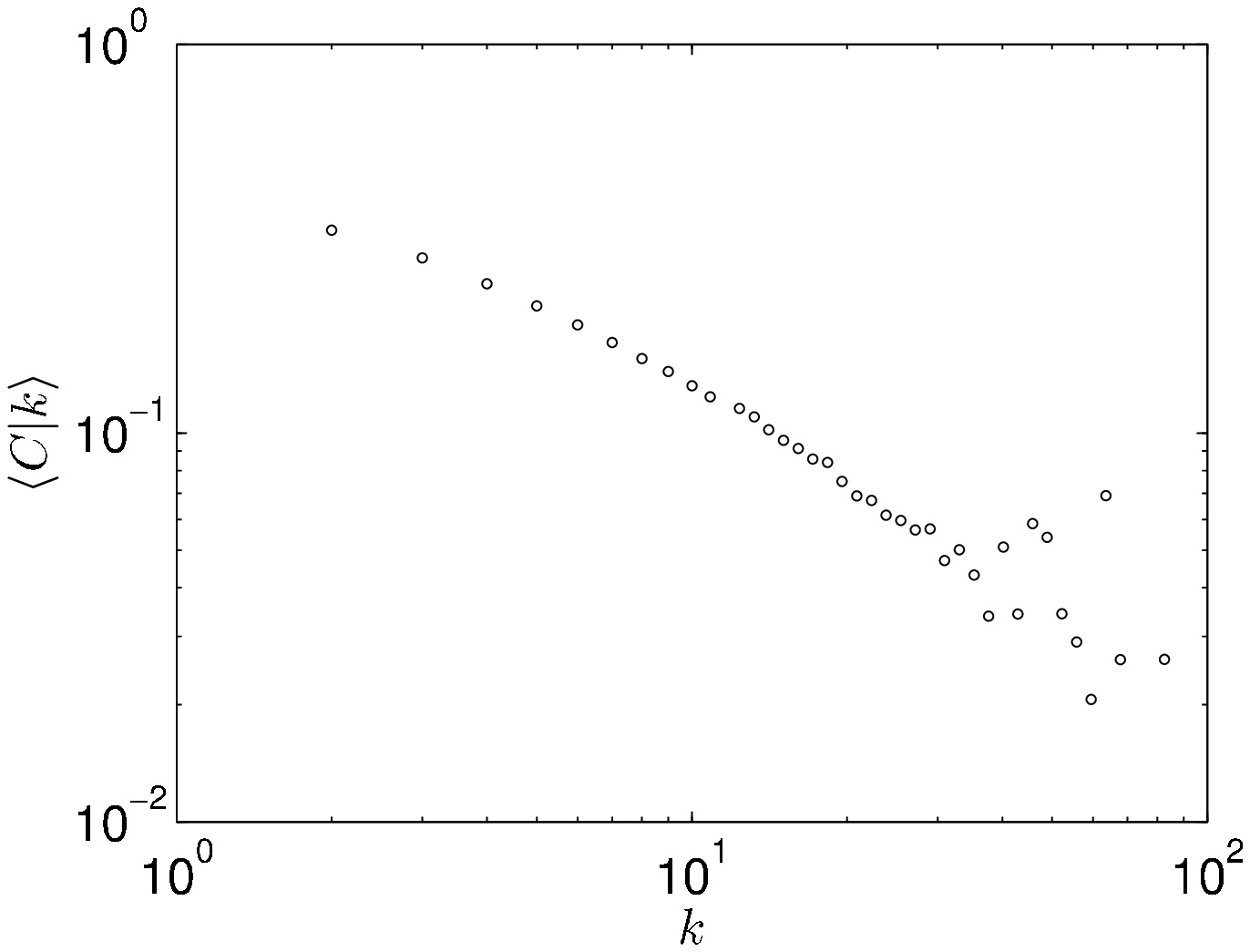}
\includegraphics[width=0.49\linewidth]{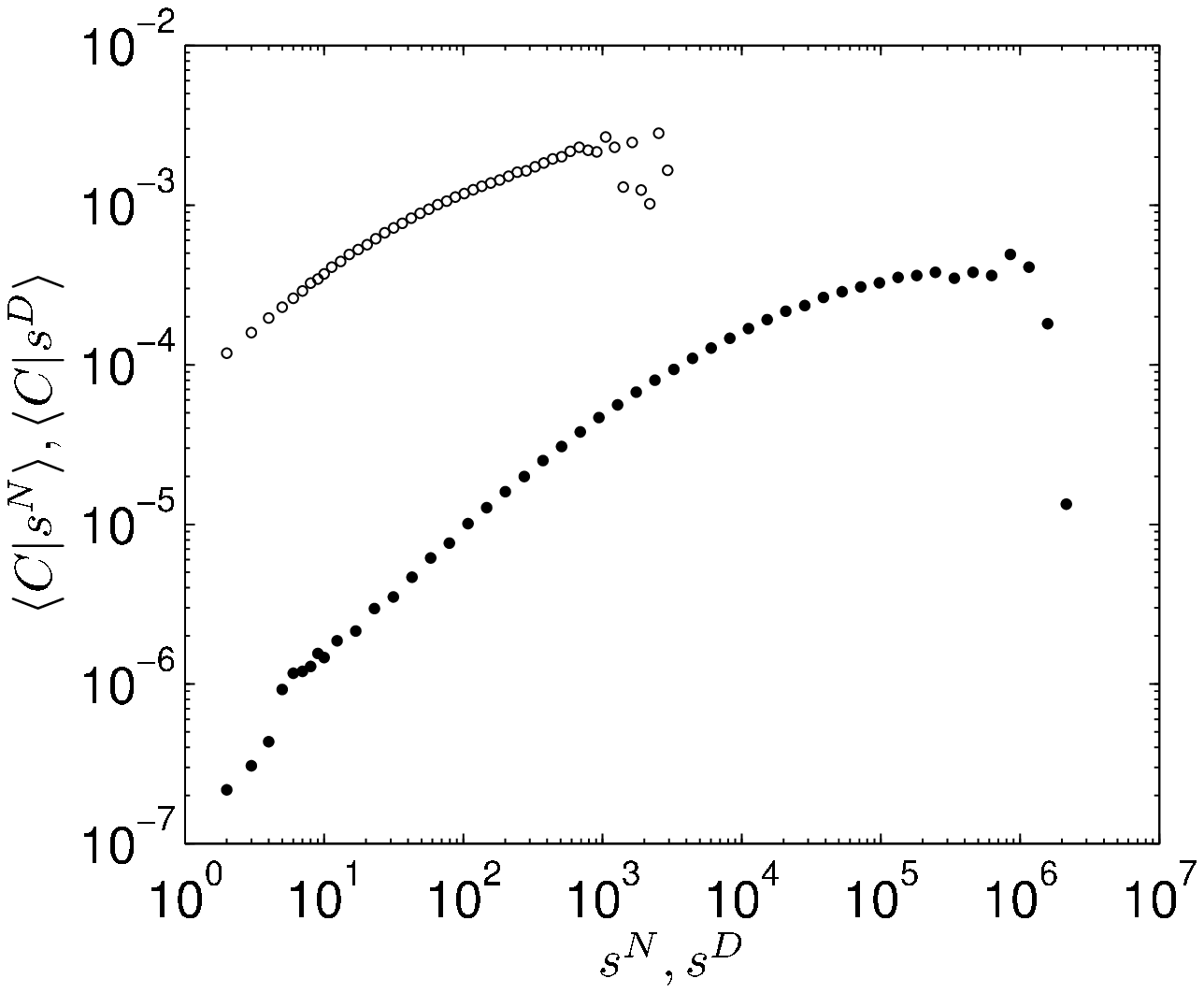}
\caption{Average (topological) clustering coefficient $\langle C|k \rangle$ (left) and average weighted clustering coefficients $\langle \tilde{C} | s^N \rangle$ and $\langle \tilde{C} | s^D \rangle$ (right) in the LCC of the mutual network. The topological clustering coefficient does not depend on weights, and is presented as a function of degree $k$ ($\circ$). In contrast, the weighted clustering coefficient is presented as a function of node strengths in terms of number of calls $s^{N}$ ($\circ$) and aggregated call duration $s^{D}$ ($\bullet$).}
\label{fig:clustering}
\end{center}
\end{figure}

The extent of clustering around a node $i$ is quantified by the (unweighted) clustering coefficient $C_i=2t_i / [k_i\left(k_i-1\right)]$, where $t_i$ denotes the number of triangles around node $i$ \cite{watts:1998}. Empirical networks have been found to have fairly high average clustering coefficients, which can be seen as manifestation of the presence of three-point correlations. Typically, one looks at the average  clustering coefficient as a function of degree $\langle C|k \rangle$, known as the clustering spectrum, as shown in Fig.~\ref{fig:clustering}. Here $\langle C|k \rangle \sim k^{-1}$ as is commonly found in many empirical networks \cite{szabo:2004}. This seems to indicate that clustering spectrum does not discriminate very well between different networks, which motivates us to adopt weighted network characteristics in Section \ref{sec:adv}.

\begin{figure}
\begin{center}
\includegraphics[width=0.49\linewidth]{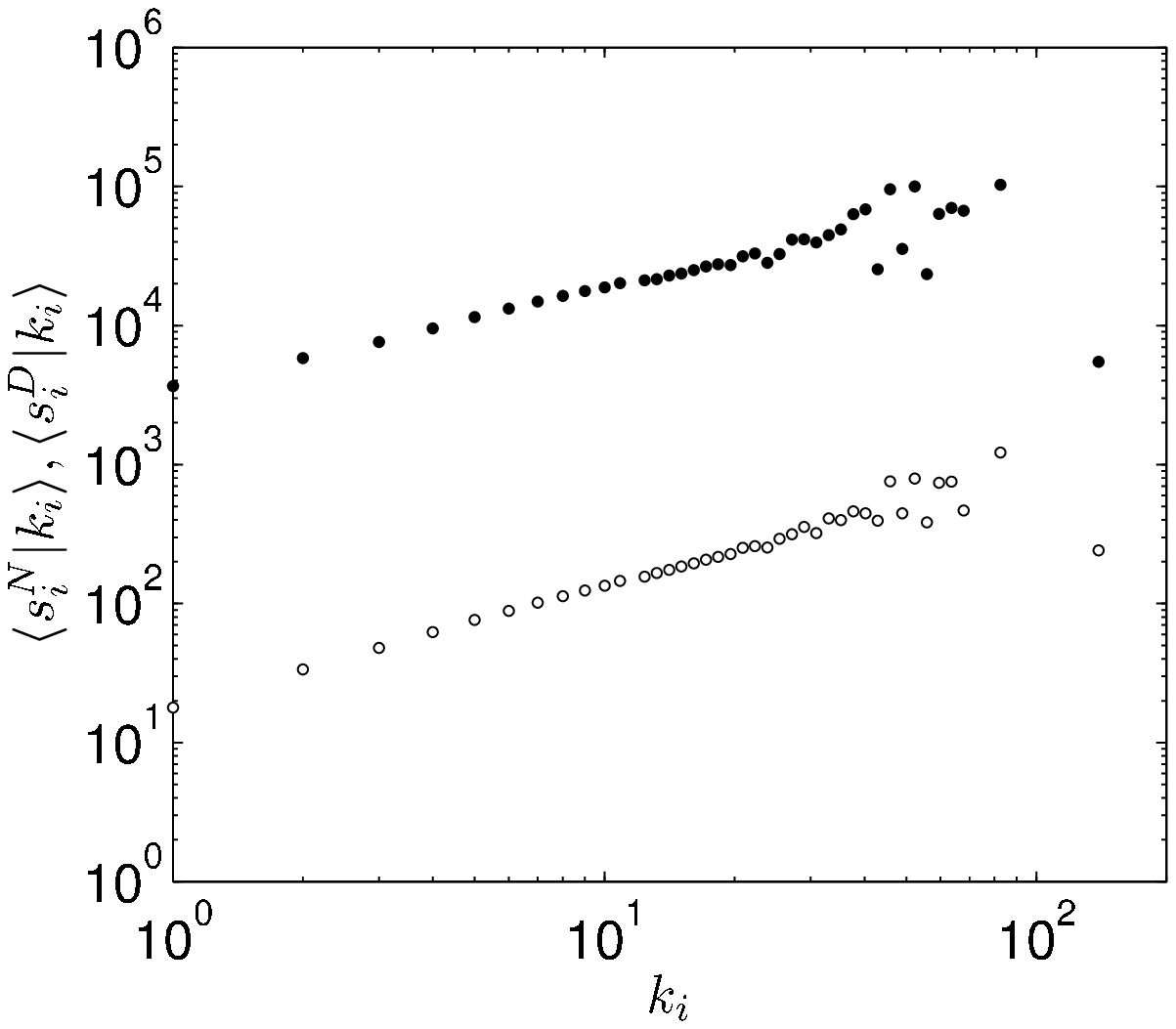}
\includegraphics[width=0.49\linewidth]{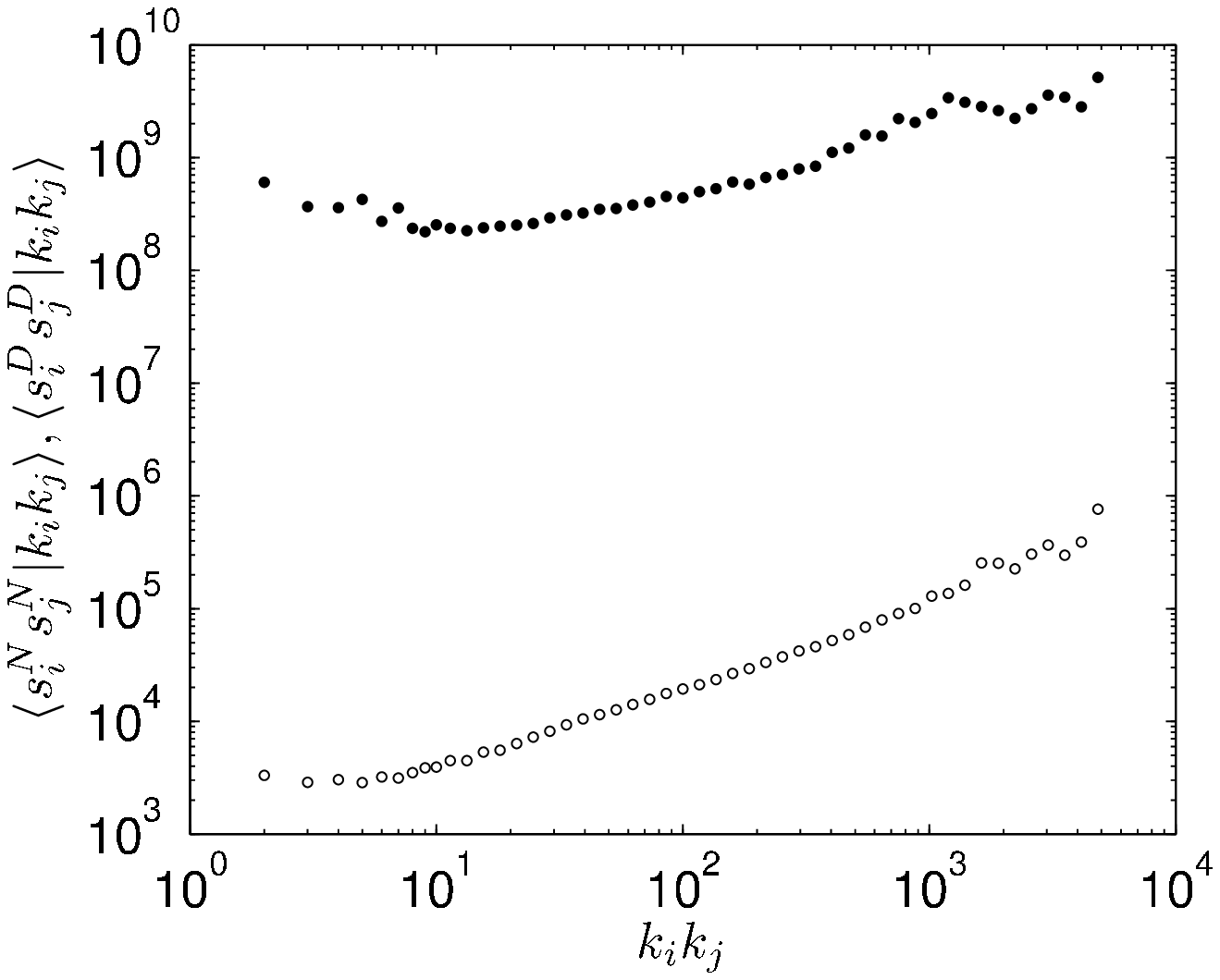}
\caption{Average strength conditional on degree in terms of the number of calls $\langle s^{N}|k \rangle$  ($\circ$) and aggregated call duration $\langle s^{D}|k \rangle$ ($\bullet$) (left) and average strength product $s_is_j$ as a function of degree product $k_ik_j$ denoted by $\langle s_i^N s_j^N | k_i k_j \rangle$ and $\langle s_i^D s_j^D | k_i k_j \rangle$.}
\label{fig:degstr}
\end{center}
\end{figure}

We have seen above that vertex degree distribution and vertex strength distribution are very similar in nature, which can be understood by examining degree-strength correlations. Average strength conditional on degree in terms of the number of calls $\langle s^{N}|k \rangle$  and aggregated call duration $\langle s^{D}|k \rangle$ are shown in Fig.~\ref{fig:degstr}. If there were no correlations between vertex degree and the weights of the links adjacent to the vertex, as can be obtained by shuffling the weights of the links, we would expect that $\langle s | k \rangle \sim k^{\alpha}$ with $\alpha = 1$, since $\langle s_i \rangle = k_i \langle w \rangle$, where $\langle w \rangle$ is the average link weight in the network. However, now we have $\langle s^D | k \rangle \sim k^{\alpha^D}$ where $\alpha^D \approx 0.8$ and $\langle s^N | k \rangle \sim k^{\alpha^N}$ where $\alpha^N \approx 0.9$, indicating that vertex strength grows somewhat more slowly than vertex degree. This is to say that individuals who talk to a large number of friends, on average, have slightly less time per friend than those who spend less time on the phone.

\begin{figure}
\begin{center}
\includegraphics[width=0.49\linewidth]{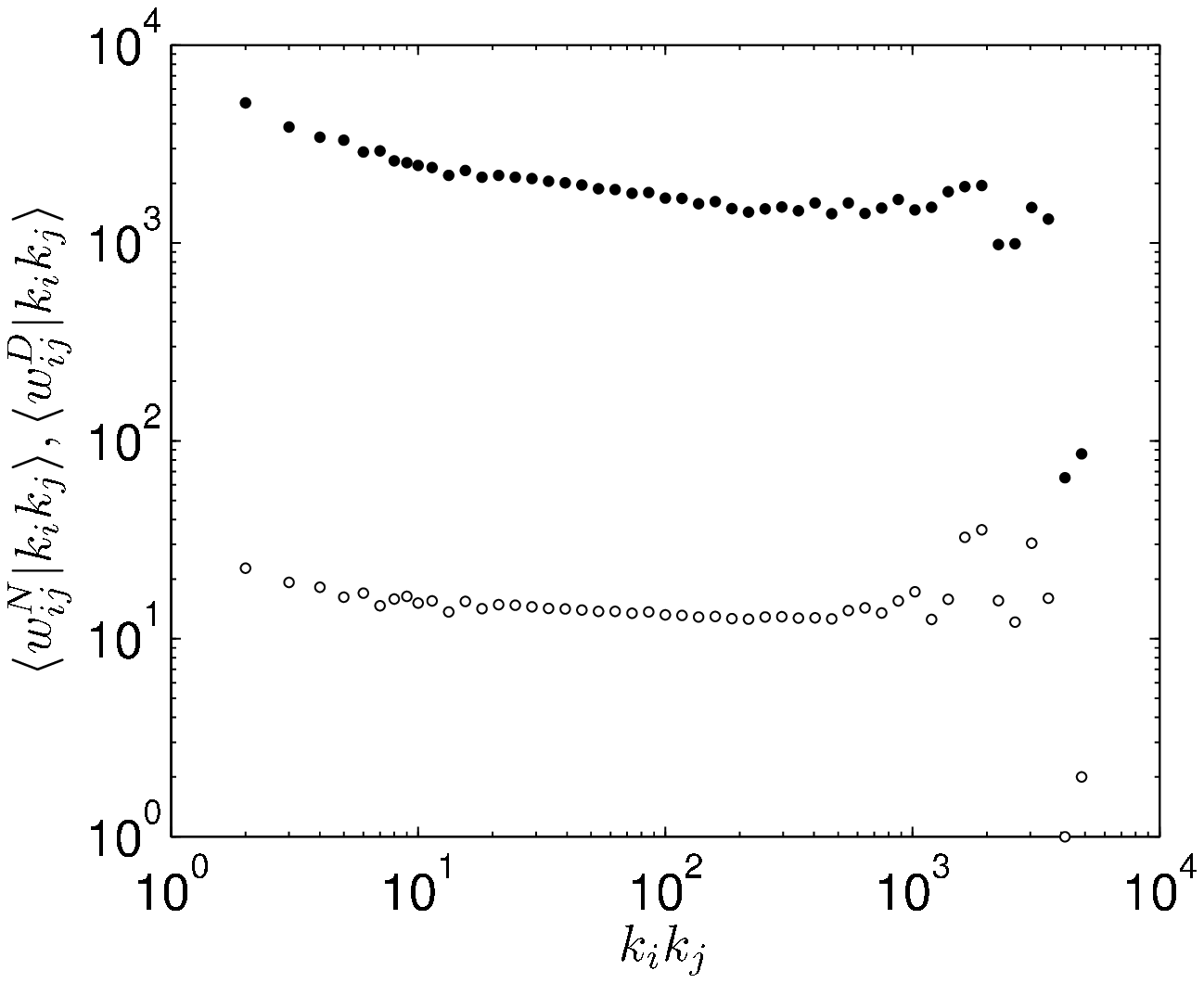}
\includegraphics[width=0.49\linewidth]{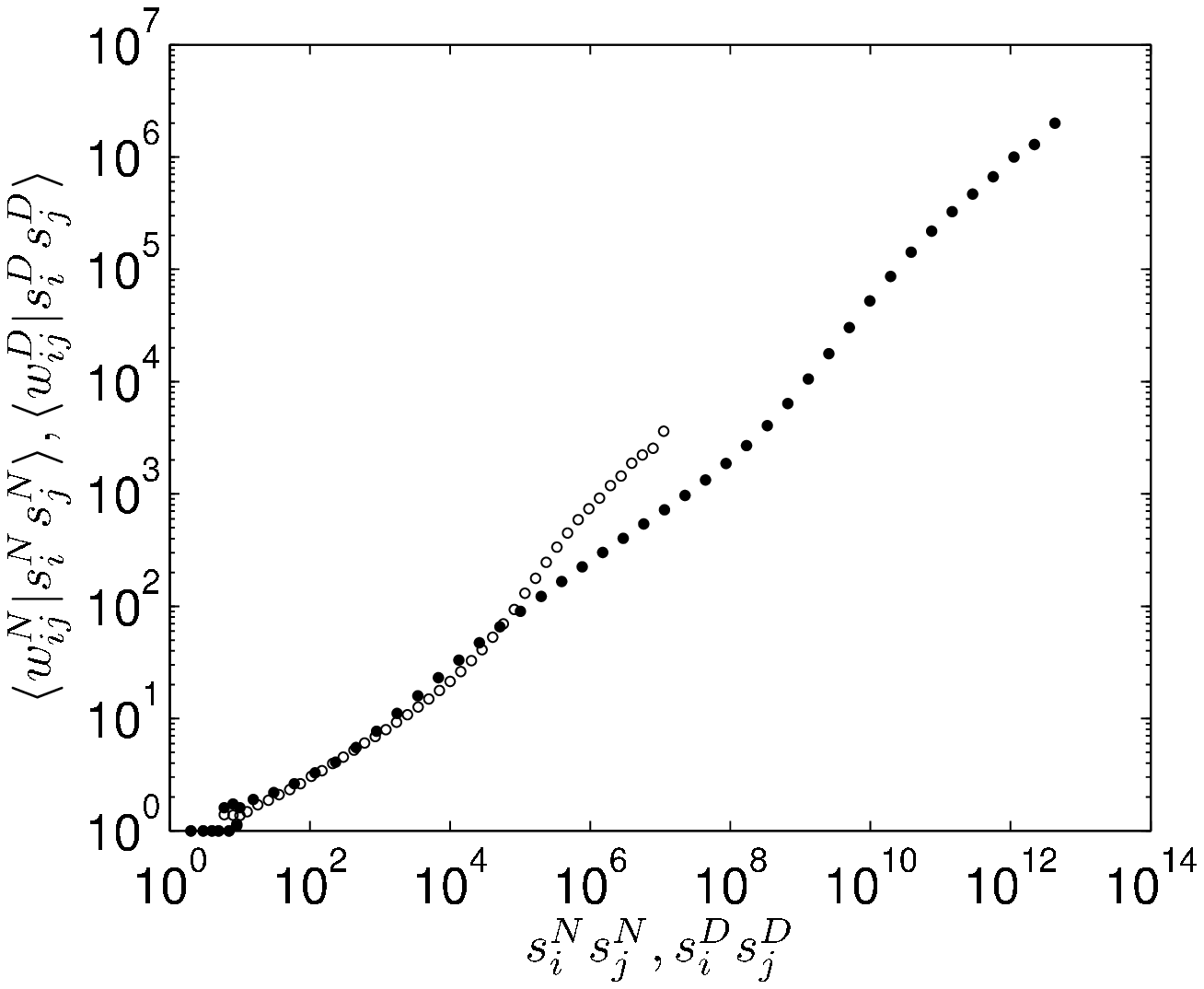}
\caption{Scaling of weights $w_{ij}^N$ ($\circ$) and $w_{ij}^D$ ($\bullet$) as a function degree product $k_ik_j$ (left) and strength product $s_is_j$ (right). }
\label{fig:weightscaling}
\end{center}
\end{figure}

We can study the strength product $s_i s_j$ as a function of degree product $k_ik_j$, the averages of which are denoted by $\langle s_i^N s_j^N | k_i k_j \rangle$ and $\langle s_i^D s_j^D | k_i k_j \rangle$, shown on the right in Fig.~\ref{fig:degstr}. In the absence of correlations, we would expect that $\langle s_i s_j | k_i k_j \rangle =  \langle w \rangle^2 \langle k_i k_j \rangle$ giving $\langle s_j^D s_j^D | k_i k_j \rangle \sim (k_i k_j)^{\beta}$ with $\beta = 1$. However, we now obtain $\beta^D \approx 0.4$ whereas $\beta^N \approx 0.7$, corresponding to sublinear growth. Let us also introduce scaling exponent for degree products such that $\langle w_{ij}^N | k_ik_j \rangle \sim (k_ik_j)^{\gamma^N}$ and $\langle w_{ij}^D | k_ik_j \rangle \sim (k_ik_j)^{\gamma^D}$ and for strength products such that $\langle w_{ij}^N | s_i^N s_j^N \rangle \sim (s_is_j)^{\delta^N}$ and $\langle w_{ij}^D | s_i^D s_j^D \rangle \sim (s_is_j)^{\delta^D}$. The plots of these quantities are shown in Fig.~\ref{fig:weightscaling}. We find that $\gamma^D \approx -0.2$ and $\gamma^N \approx -0.1$, indicating that the links weights, wheter measured in terms of $w_{ij}^D$ or $w_{ij}^N$, are practically independent of the degree product $k_ik_j$. This shows that links weights are not determined by the absolute number of friends (node degrees) of $v_i$ and $v_j$. In contrast, as we will see in Section \ref{sec:link}, link weights are dependent on the relative proportion of common neighbors (link overlap). For the latter exponents we have $\delta^N \approx \delta^D \approx 0.5$, such that $w_{ij}$ scales as the geometric mean of the strengths of the adjacent nodes.

Putting these structural properties together, we have seen that the network has a very steep degree distribution, resulting in few highly connected nodes, and even they are not as connected as hubs in scale-free networks are. The two weights, number of calls and aggregate call duration, are strongly correlated, and both yield steep strength distributions for nodes. This can be understood in light of the only slightly sublinear dependence of strength on degree, governed by the exponent $\alpha$. Topologically the network is assortative, but not weight-assortative for a large majority of nodes. The weight of a given link is almost independent of the product of the degrees of adjacent nodes as governed by the almost vanishing exponent $\gamma$, but depends on the geometric mean of the strengths of the adjacent nodes as indicated by the exponent $\delta$. 

\section{Advanced network characteristics}
\label{sec:adv}

Study of purely topological properties of networks, as was done in Section \ref{sec:basic}, is a useful starting point, but incorporating weights in the analysis is important, as it can enhance our understanding of the structural properties of the network. This motivates us to proceed to weighted network characteristics. Here important concepts are subgraph intensity and subgraph coherence that can be used to study the coupling between network structure and interaction strengths \cite{onnela:2005}. The intensity of subgraph $g$ with vertices $v_g$ and links $\ell_g$ is given by the geometric mean of its weights as

\begin{equation} 
i(g)=\left(\prod_{(ij)\in \ell_g} w_{ij}\right) ^{1/|\ell _g|}, 
\label{eq:geom_mean} 
\end{equation} 

\noindent where $|\ell _g|$ is the number of links in $\ell_g$ \cite{onnela:2005}. Note that the unit of intensity is the same as the unit of network weights. To characterize the homogeneity of weights in a subgraph, we defined subgraph coherence $q(g)$ as the ratio of the geometric to the arithmetic mean of the weights as
\begin{equation} 
q(g) = i(g) |\ell _g|/\sum_{(ij)\in \ell_g} w_{ij}.
\label{eq:coherence} 
\end{equation} 

\noindent Here $ q(g) \in [0,1]$ and it is close to unity only if the weights of subgraph $g$ do not differ much, i.e. are internally coherent \cite{onnela:2005}.

\begin{figure}
\begin{center}
\includegraphics[width=0.49\linewidth]{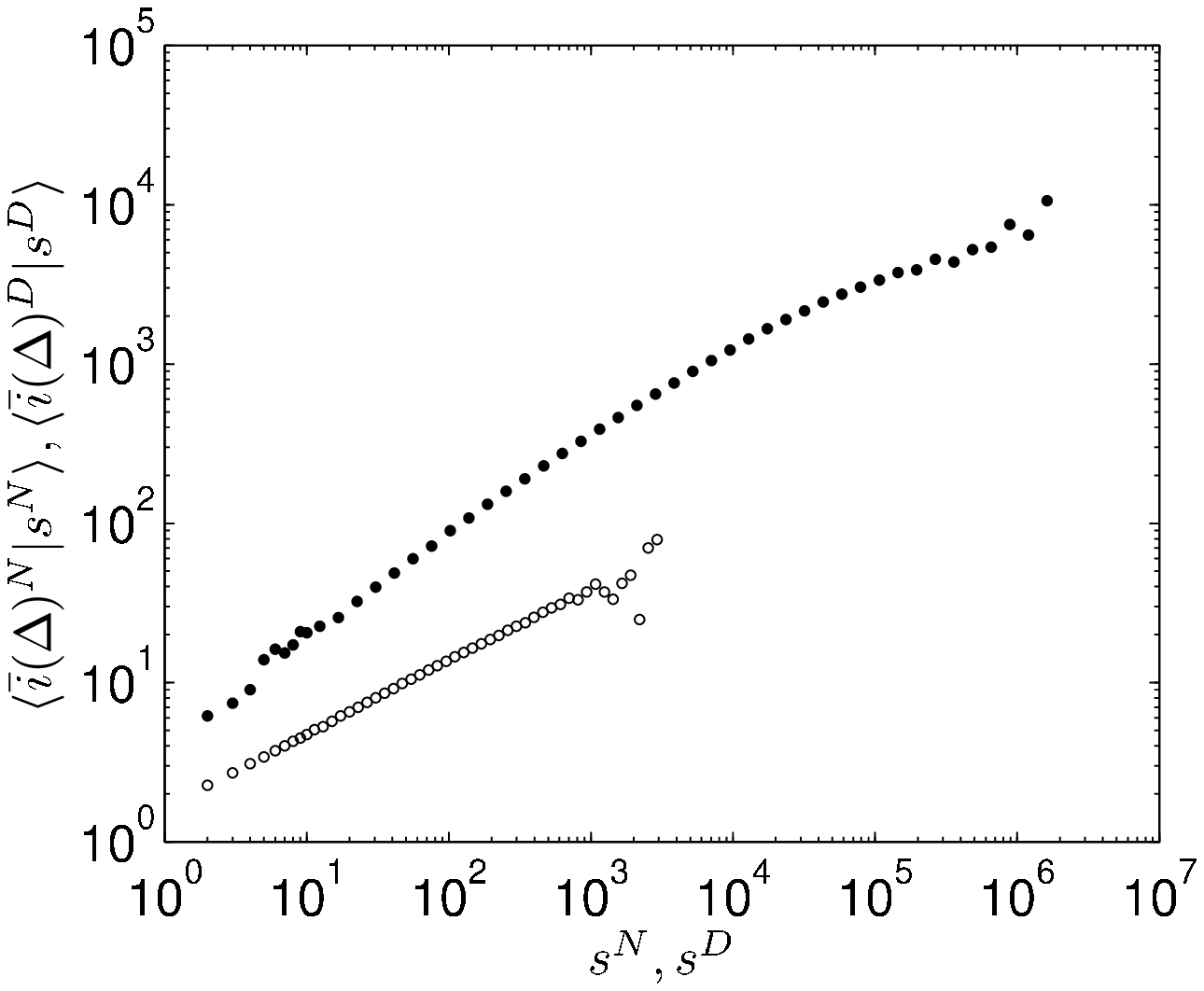}
\includegraphics[width=0.49\linewidth]{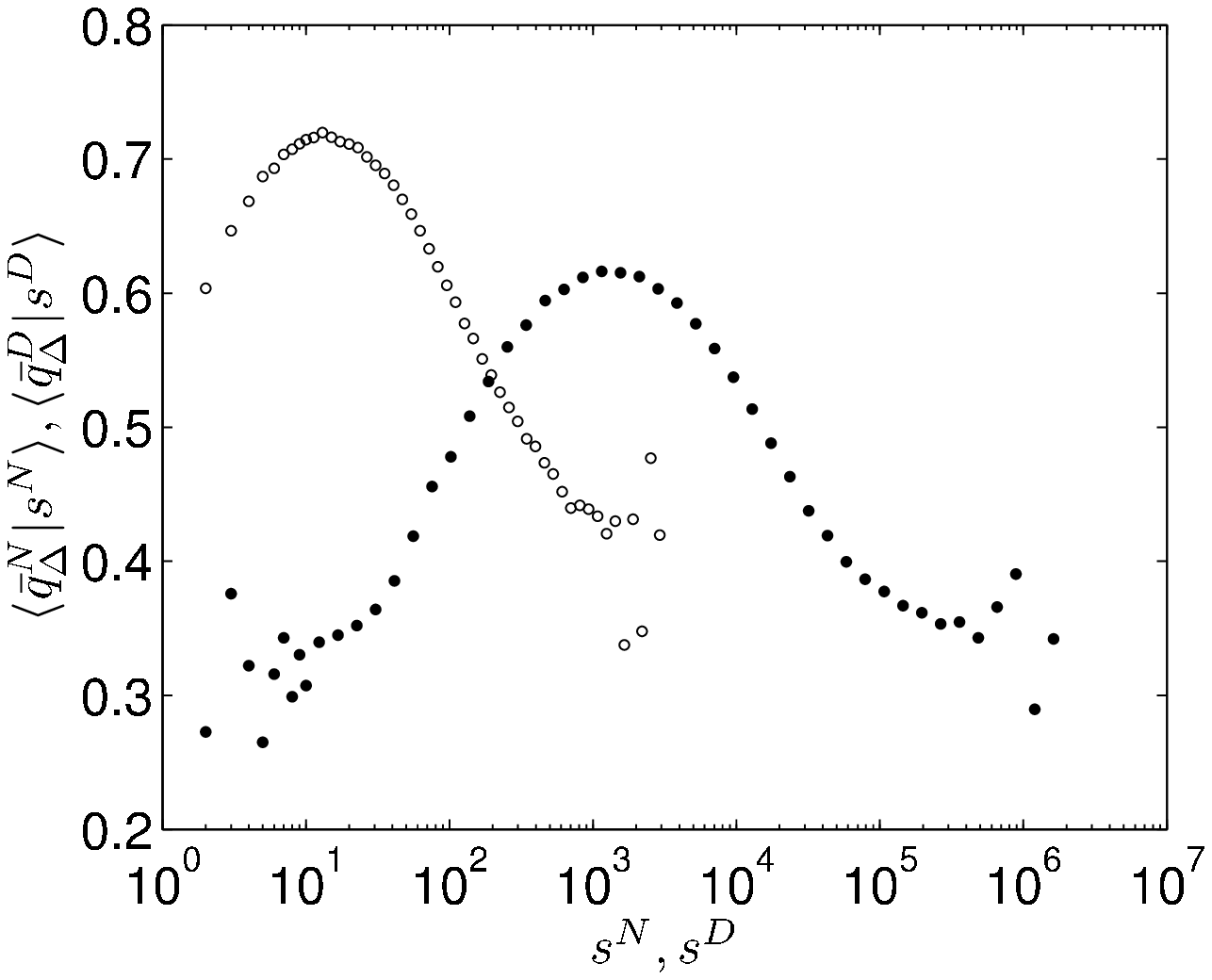}
\caption{Average mean intensity of triangles $\langle \bar i(\Delta)^N | s^N\rangle$ and $\langle \bar i(\Delta)^D | s^D\rangle$(left), and average mean coherence $\langle \bar q(\Delta)^N | s^N\rangle$ and $\langle \bar q(\Delta)^D | s^D\rangle$ (right) as a function of node strengths $s^N$ ($\circ$) and $s^D$ ($\bullet$).}
\label{fig:intcoh}
\end{center}
\end{figure}

The average intensity of subgraph $g$ at node $k$ is given by  $\bar i_k(g) = (1/t_k) \sum_{g_k} i(g)$, where $\sum_{g_k}$ denotes a sum over all topologically equivalent subgraphs containing node $k$. We can average this over all nodes that participate in one instance of the subgraph, denoted by $\langle \bar i(g) \rangle = (n(g) |v_g|)^{-1} \sum_{k} \bar i_k (g)$, where $n(g)$ is the number of subgraphs $g$ in the network and $ |v_g|$ is the number of nodes in subgraph $g$. Regarding notation, we emphasize that $\bar i_{k}(\Delta)$ denotes the mean intensity of triangles around a particular node $k$, where the mean is taken over all triangles attached to the node, whereas $\langle \bar i(\Delta) | s\rangle$ denotes average taken over all nodes whose strength is approximately $s$. The behavior of average intensity of triangles as a function of node strength, $\langle \bar i(\Delta)^N | s^N\rangle$ and $\langle \bar i(\Delta)^D | s^D\rangle$, and average mean coherence, $\langle \bar q(\Delta)^N | s^N \rangle$ and $\langle  \bar q(\Delta)^D | s^D \rangle$, are shown in Fig.~\ref{fig:intcoh}. We find that $\langle \bar i(\Delta)^N | s^N\rangle \sim (s^N)^{\epsilon^{N}}$, where $\epsilon^{N} \approx 0.5$ and $\langle \bar i(\Delta)^D | s^D\rangle \sim (s^D)^{\epsilon^{D}}$, where $\epsilon^{D} \approx 0.7$. The behavior of average mean coherence $\langle \bar q(\Delta)^D | s^D\rangle$ is markedly different from that of the intensity, achieving a  maximum at $s^{D} \approx 10^3$.

To consider the effect of weights on the clustering properties of the network, we adopt the definition proposed for a weighted clustering coefficient in \cite{onnela:2005}, leading to 

\begin{equation}
\tilde{C}_i=\frac{1}{k_i\left(k_i-1\right)} \sum_{j,k}\left(\hat{w}_{ij}\hat{w}_{ik}\hat{w}_{jk}\right)^{1/3} = C_i \bar i_i(\triangle), 
\label{eq:wc}
\end{equation}

\noindent where $\bar i_i(\triangle)$ denotes the average intensity of triangles at node $i$. The weights are normalized by the maximum weight in the network, $\hat{w}_{ij}=w_{ij}/\max(w)$, required for reasons of compatibility with the topological clustering coefficient, and the contribution of each triangle depends on all of its edge weights \cite{saramaki:2005,saramaki:2007}. Note that the weighted clustering coefficient can be written as the product of the unweighted clustering coefficient and the average intensity of triangles at a node as shown in Eq. ~\ref{eq:wc}. Thus triangles in which each edge weight equals $\max(w)$ contribute unity to the sum, while a triangle having one link with a negligible weight will have a negligible contribution to the clustering coefficient. Results are shown in Fig.~\ref{fig:clustering} next to the unweighted (topological) clustering coefficient. It is clear that the behavior for number of calls and aggregate duration is very similar. For the duration we assume again that a crossover sets in at $s_x^D \approx 10^4$. Up-to this point the power law $\langle \tilde{C} | s^D \rangle \sim (s^D)^{\zeta^D}$ with $\zeta^D \approx 0.8$ gives an acceptable fit. However, the behavior of $\langle \tilde{C} | s^N \rangle$ cannot really be described by a power-law. 

The local structure of unweighted networks can be characterized by the appearance of small subgraphs, which have been related to the functionality of several networks \cite{milo:2004,milo:2002}. This is done by studying the number of times a subgraph of interest appears in the network, but to draw statistical conclusions about the appearance frequency of subgraphs, a reference system needs to be specified, which can be seen as analogous to setting up a null hypothesis $H_0$ in the statistics literature. The reference system is usually established by rewiring the network while conserving its degree distribution in order to remove local structural correlations present in the original network. Statistical significance of motifs is usually measured in terms of a $z$-score statistic \cite{milo:2002}. Here we have chosen just to provide the number of fully connected subgraphs up-to order $k=10$ in Table \ref{table:counts} for both the empirical network and a corresponding Erd\"{o}s-R\'enyi network \cite{erdos:1960}.

\begin{table}
\begin{center}
\begin{tabular}{| l | l  | l |}
\hline
Order & Empirical count & ER expectation \\
\hline
1 & $6.3 \times 10^6$ & $ 6.3 \times 10^6$ \\
2 & $17 \times 10^6$ & $17 \times 10^6$ \\
3 & $5.6 \times 10^{6}$ & $2.6 \times 10^1$\\
4 & $1.4 \times 10^{6}$ & $2.5 \times 10^{-11}$ \\
5 & $2.7 \times 10^{5}$ & $1.7 \times 10^{-29}$ \\
6 & $4.5 \times 10^{4}$ & $7.8 \times 10^{-54}$ \\
7 & $6.8 \times 10^{3}$ & $2.7 \times 10^{-84}$ \\
8 & 799 & $7.0 \times 10^{-121}$ \\
9 & 61 & $1.4 \times 10^{-163}$  \\
10 & 2 & $2.0 \times 10^{-212}$\\
\hline
\end{tabular}
\caption{Number of cliques of order $k=1, 2, \ldots, 10$ in the empirical network (Empirical count) and their expectation values in a corresponding Erd\"{o}s-R\'enyi network (ER expectation) \cite{erdos:1960}. Note that $k=1$ corresponds to the number of nodes $N= 6282226 $ and $k=2$ to the number of links $L= 16828910$, which are the same in the empirical and random network. These values of $N$ and $L$ give the link formation probability in the ER graph as $p = 2L[N(N-1)] \approx 8.5 \times 10^{-7}$. The expected number $E[X]$ of subgraphs with $k$ nodes and $\ell$ links is given by $E[X] = {N \choose k} (k!/a) p^{\ell}$, where $\ell = k(k-1)/2$ and $a = k!$ is the number of graphs that are isomorphic to one another, i.e., automorphic, defined as adjacency-preserving permutation of the vertices of the graph \cite{bollobas:2001}. Here the empirical network is a non-mutual one formed from the aggregated calls of 12 weeks. Note that subgraphs are counted multiple times, such that one subgraph of order $k$ contains $k$ subgraphs of order $k-1$ and so on. For example, one subgraph with $k=5$ will also be counted as five instances of subgraph of order $k=4$, and $5 \times 4 = 20$ instances of subgraph of order $k=3$. The presence of high-order topological correlations, as manifest by the existence of cliques beyond order three (triangles) in the empirical network, makes is starkly different from an ER graph, in which high-order cliques have astronomically low probability to be present.}
\label{table:counts}
\end{center}
\end{table}

\begin{figure}
\begin{center}
\includegraphics[width=0.8\linewidth]{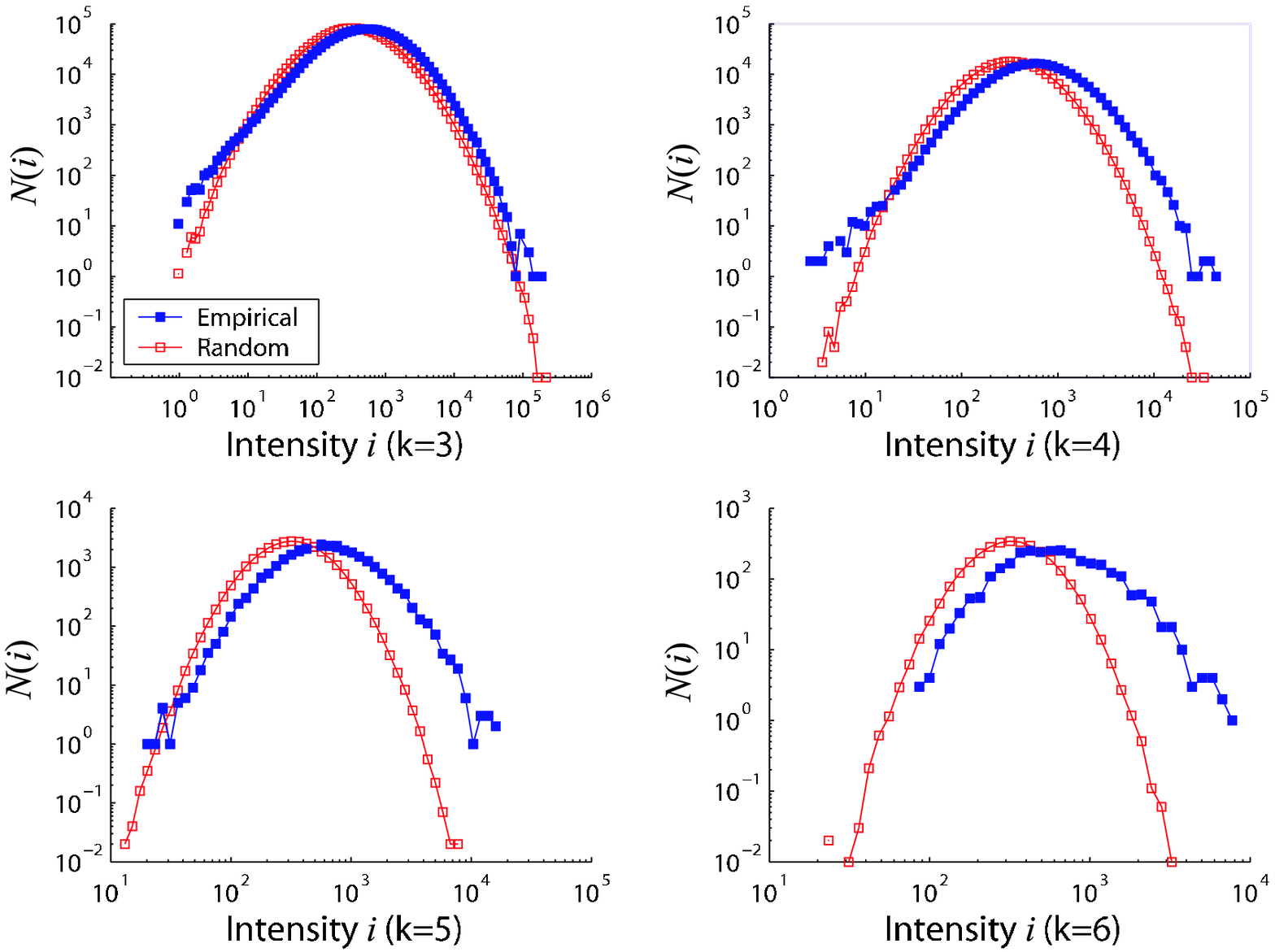}
\caption{Distribution of subgraph intensity based on aggregate call duration weights $w_{ij}^D$ for cliques $g$ of order $k=3,4,5,6$ in the LCC of the empirical mutual network (solid blue squares) and in a reference ensemble (open red squares). Number of subgraphs of intensity $i$ in the empirical network is given by $n^E (g,i)$ and their average number in 100 realizations of the ensemble by $\bar n^R (g,i)$. A realization of the ensemble is obtained by shuffling the weights $w_{ij}^D$ in the empirical network while keeping its topology fixed. Note that both horizontal and vertical scales vary between the panels.}
\label{fig:subintensity}
\end{center}
\end{figure}

The motif framework has been generalized to weighted networks \cite{onnela:2005}, with the motivation of studying the nature of coupling between interactions strengths (link weights $w_{ij}$) and local network topology (an ensemble of subgraphs $g$). We set up a \emph{weight permuted reference} by simply shuffling the weights in the network, which removes weight correlations while leaving the network topology unaltered. Any deviation in motif intensities between the empirical and reference system has a straightforward interpretation: the local organisation of weights in the empirical network is not random. While the $z$-score may be generalized to weighted networks as demonstrated in \cite{onnela:2005}, it has the same shortcoming as the $z$-score has for unweighted networks, namely, that it is based on just one number characterizing the empirical network and two numbers characterising the reference distribution. We follow an alternative approach here introduced in \cite{onnela:2006c}, which makes use of the entire intensity distribution $P^E(g)$ for subgraphs $g$ in the empirical network to the intensity distribution $P^R(g)$ in the corresponding reference ensemble. Now the problem becomes one of comparing two distributions with one another for which several tools are available, such as the standard Kolmogorov-Smirnov test or the Kullback-Leibler divergence \cite{kullback:1951}. This approach suggests a shift in perspective from regarding subgraphs as discrete objects that either exist or not to a continuum of subgraph intensities and coherences.

\begin{figure}
\begin{center}
\includegraphics[width=0.8\linewidth]{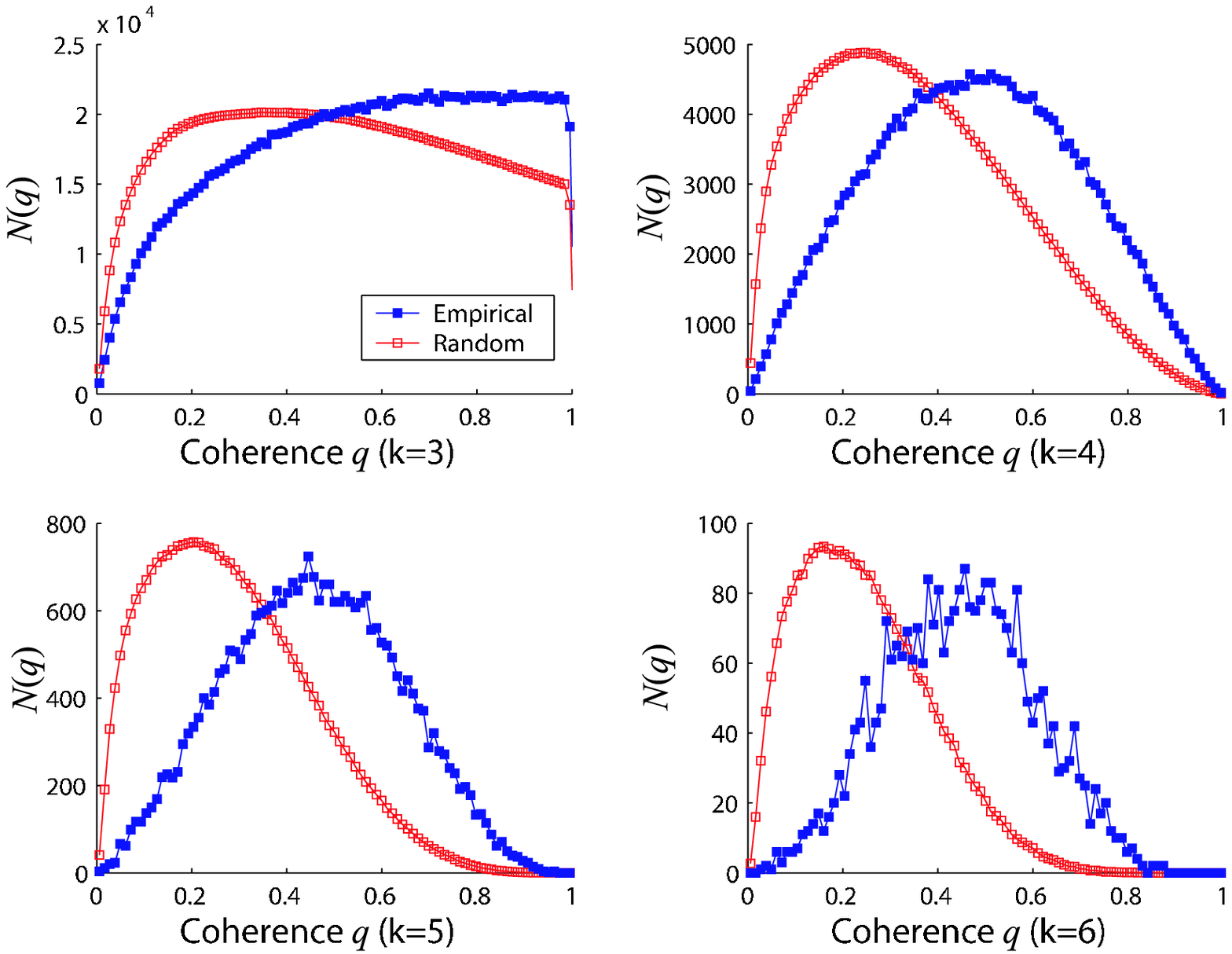}
\caption{Distribution of subgraph coherence based on aggregate call duration weights $w_{ij}^D$ for cliques of order $k=3,4,5,6$ in the LCC of the empirical mutual network $n^E(g,q)$ (solid blue squares) and in the reference ensemble $\bar n^R(g,q)$ (open red squares). Note that both horizontal and vertical scales vary between the panels.}
\label{fig:subcoherence}
\end{center}
\end{figure}

Results are shown for intensity in Fig.~\ref{fig:subintensity} and for coherence in Fig.~\ref{fig:subcoherence}. Comparing the subgraph intensity distribution shows that the empirical subgraphs have considerably higher intensities than their random counterparts. Noting in particular the vertical logarithmic scale, we see that some high intensity subgraphs can be 10-1000 times more frequent in the empirical than in the reference ensemble. Especially for the larger subgraphs, e.g. $k=6$, there are some extremely high intensity subgraphs in the empirical network, which are never created randomly in the reference ensemble. Similarly, the subgraphs in the empirical network are more coherent than their randomized counterparts. The differences become larger as we move to more complex subgraphs, the reason being that it is increasingly unlikely to create coherent subgraphs with many links by chance. Putting the results on intensity and coherence together, link weights within cliques are higher and more similar in magnitude that expected in a randomized reference system. Consequently, there are important correlations between local network structure at the level of cliques, or communities, and interactions strengths within them.

\section{Single link properties}
\label{sec:link}

Let us now move from subgraphs to study the properties of links and their immediate neighborhood. We quantify the topological overlap of the neighborhood of two connected nodes $i$ and $j$ by the relative overlap of their common neighbors, defined as

\begin{equation}
O_{ij} = \frac{n_{ij}}{(k_i-1) + (k_j-1) - n_{ij}}, 
\end{equation}

\noindent where $n_{ij}$ is the number of neighbors common to both nodes $i$ and $j$ \cite{onnela:2007}. It is worth pointing out that this is similar, but not identical, to the edge-clustering coefficient as introduced by Radicchi \emph{et al.} as 

\begin{equation}
C_{ij} = \frac{n_{ij}}{\min(k_i, k_j) - 1}, 
\label{eq:linkclust}
\end{equation}

\noindent where $\min(k_i, k_j) - 1$ is the maximum possible number of triangles around the $(i,j)$ edge  \cite{radicchi:2004a}. Edge-clustering coefficient reflects the probability that a pair of connected vertices has a common neighbor, whereas overlap is the fraction of common neighbors a pair of connected vertices has. The reason for using $O_{ij}$ as opposed to $C_{ij}$ is that the denominator of Eq.~\ref{eq:linkclust} gives rise to two undesirable features in the context of social networks. First, consider a subgraph in which vertices $i$ and $j$ are connected only with a single link such that $k_i = 1$ and $k_j > 1$, where vertex $i$ is a leaf of the network. We now have $O_{ij} = 0$ indicating that these two individuals have no common friends, which seems a reasonable conclusion, whereas $C_{ij}$ is either not defined or diverges as the denominator tends to zero. Second, consider a triangle $(i,j,k)$ such that $k_i = 2$, $n_{ij} = 1$ and $k_j \ge 2$. If $k_j=2$, then both $O_{ij} = 1$ and $C_{ij} = 1$. However, if $k_j > 2$, we still have $C_{ij} = 1$ for all values of $k_j$, whereas $O_{ij} = 1 / (k_j - 1)$. This is to say that the overlap of common friends decreases as $k_j$ increases since, although $i$ and $j$ still have just one common friend ($n_{ij}=1$), the overlap of their common friends decreases as vertex $j$ acquires new friends ($k_j$ increases). This is a reasonable feature of an overlap measure in a social context. Finally, as a general remark, since the overlap is a property of the link, it has the desirable property that, unlike $C_{ij}$, it is symmetric  with respect to its arguments $k_i$ and $k_j$.

\begin{figure}
\begin{center}
\includegraphics[width=0.49\linewidth]{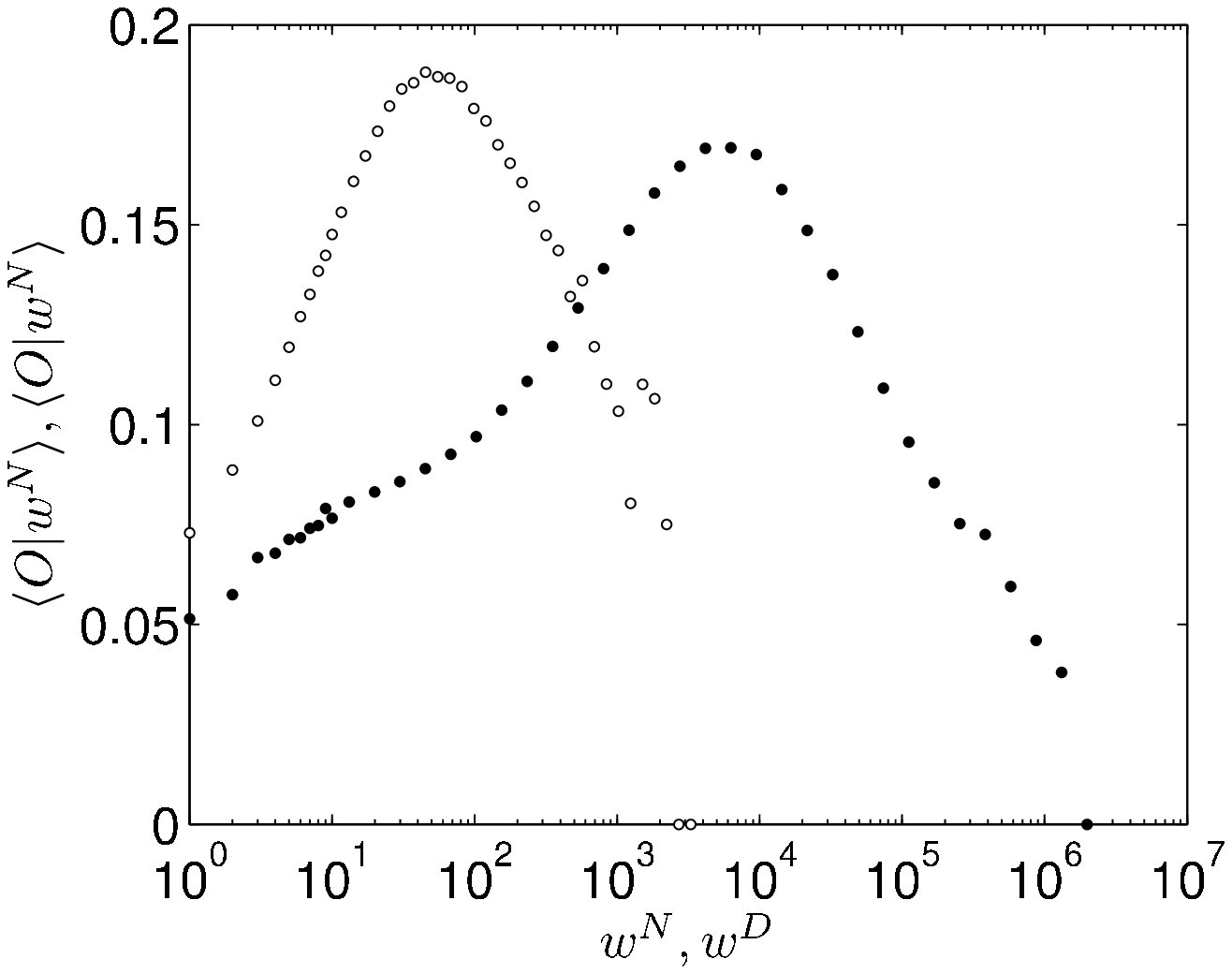}
\includegraphics[width=0.49\linewidth]{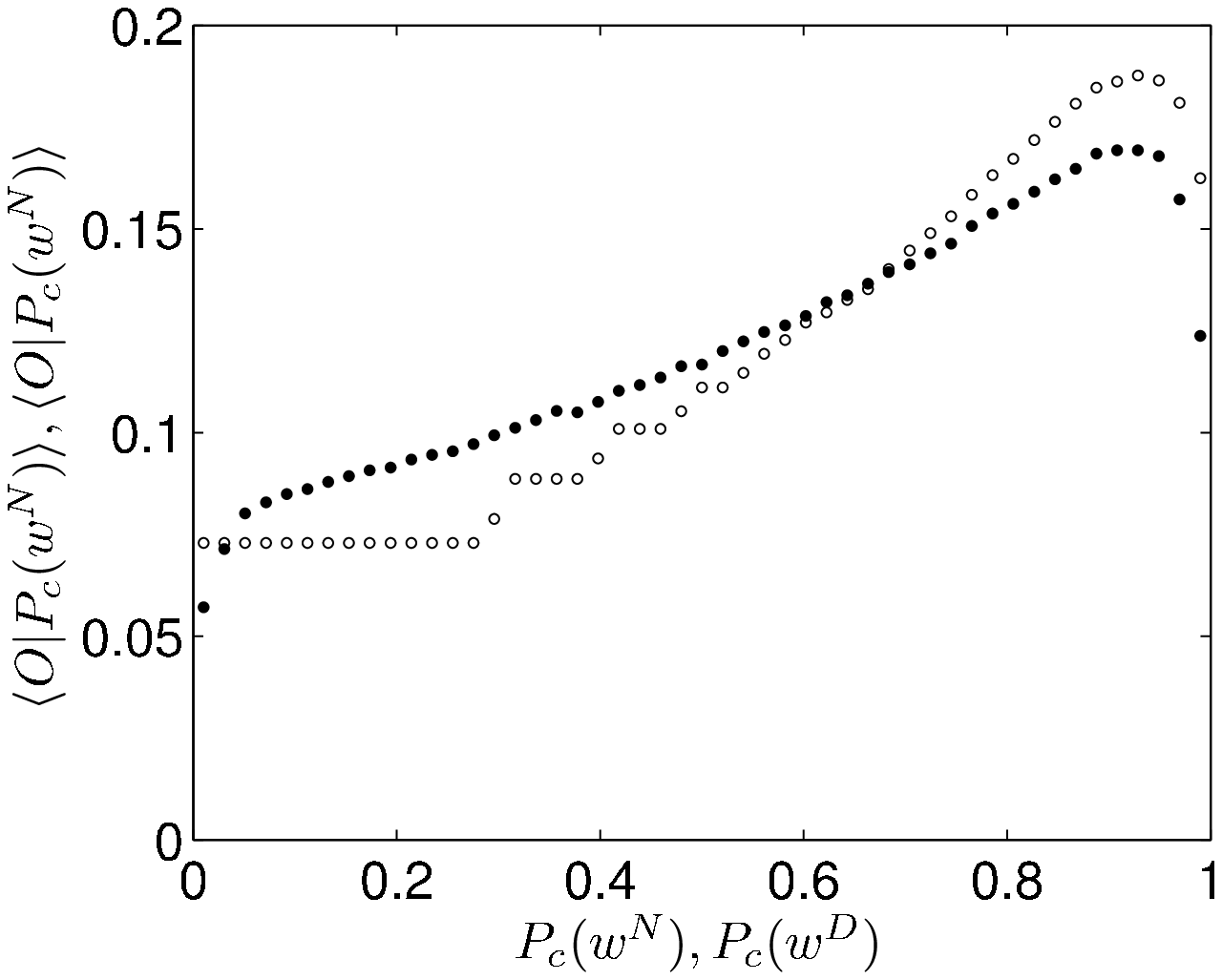}
\caption{Average overlap as a function of absolute link weight $\langle O|w^D \rangle$ (left) and cumulative link weight $\langle O|P_{c}(w) \rangle$ (right) for $w^N$ ($\circ$) and $w^D$ ($\bullet$). }
\label{fig:overlap}
\end{center}
\end{figure}

The behaviour of average overlap as a function of absolute link weight $\langle O|w^D \rangle$ and cumulative link weight $\langle O|P_{c}(w) \rangle$ is shown in Fig.~\ref{fig:overlap}. The cumulative link weight is defined in the following way. Let $P_{<}(x) = \int_{-\infty}^{x} p(w) \, dw$, where $p(w)$ is the probability density function for the link weights (either $w^N$ or $w^D$). We define $P_{c}(w) = \psi \in [0,1]$ if $P_<^{-1} (\psi - \Delta \psi) < w \le P_<^{-1} (\psi + \Delta \psi) $, where $P_<^{-1} (\cdot)$ is the inverse cumulative density function of link weights and $\Delta \psi=1/50$. Average overlap $\langle O|w^D \rangle$ increases upto $s^D \approx 10^4$, after which it declines strongly. However,  $\langle O|P_{c}(w) \rangle$ shows that the declining trend is applicable to only some 5\% of links, resulting from these individuals communicating predominantly just one other person as explored in \cite{onnela:2007}. Note that $s_x^D \approx 10^4$ was the crossover point in the distribution of $\langle s_{nn}^D | s^D \rangle$ and $\langle \tilde{C} | s^D \rangle$, indicating that the behavior of these high-strength nodes is different from that of the rest. The high strength of these nodes derives from the top 5\% of heavy links that also behave in an anomalous way as discussed in detail in \cite{onnela:2007}.

Could the result concerning overlap $O_{ij }$ vs. link weight $w_{ij}$ be affected by the fact that the phone call data is from a single operator and, consequently, calls to phone subscriptions managed by other operators are not included? Let us assume that an individual in the population has a probability $p = 0.2$ of having a subscription governed by the operator the data comes from. We assume that the nodes are all identical and that the probability of a node being governed by the operator is independent of the probability of its neighbor being governed by the operator. Given these assumptions, we can interpret $p$ as the probability of a randomly chosen node being governed by the operator and, consequently, its being included in our network. Consequently, the probability for a link to be included in the network is $p^2$ and that for a triangle is $p^3$. These probabilities give rise to expected number of nodes, links, and triangles $\hat N = N / p = 5N$, $\hat L = L / p^2 = 25L$, and $\hat T = T / p^3 = 125T$, respectively. These numbers indicate that the expected number of links and triangles in the underlying (unobserved) network, to which we have only partial visibility by virtue of having a one-operator sample of it, are 25 times the number of links and 125 times the number of triangles in the observed network, respectively. Since the value of $p$ affects the number of observed nodes, links, and triangles in the sample, it is important to consider how it may affect overlap $O_{ij}$.

\begin{figure}
\begin{center}
\includegraphics[width=0.49\linewidth]{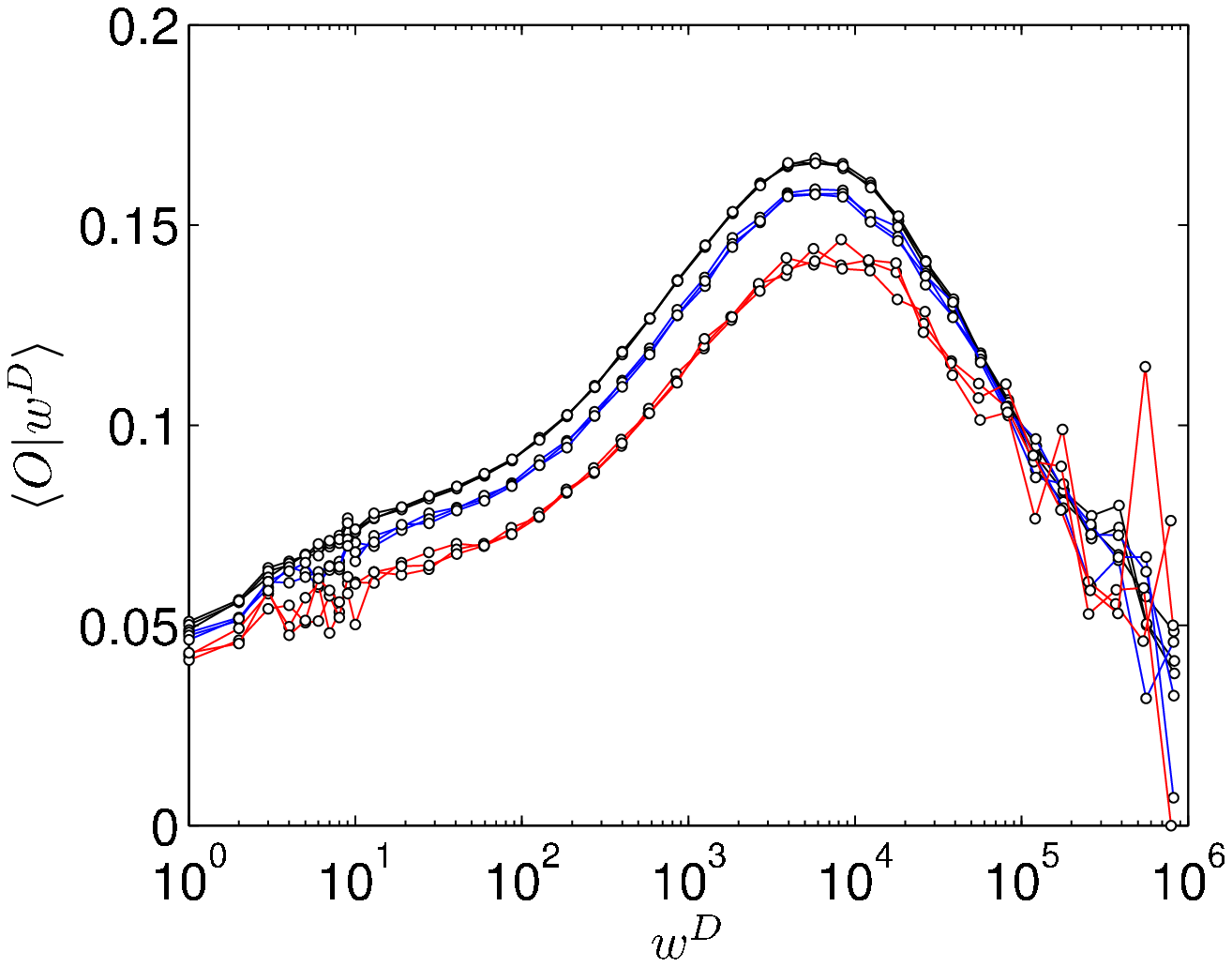}
\includegraphics[width=0.49\linewidth]{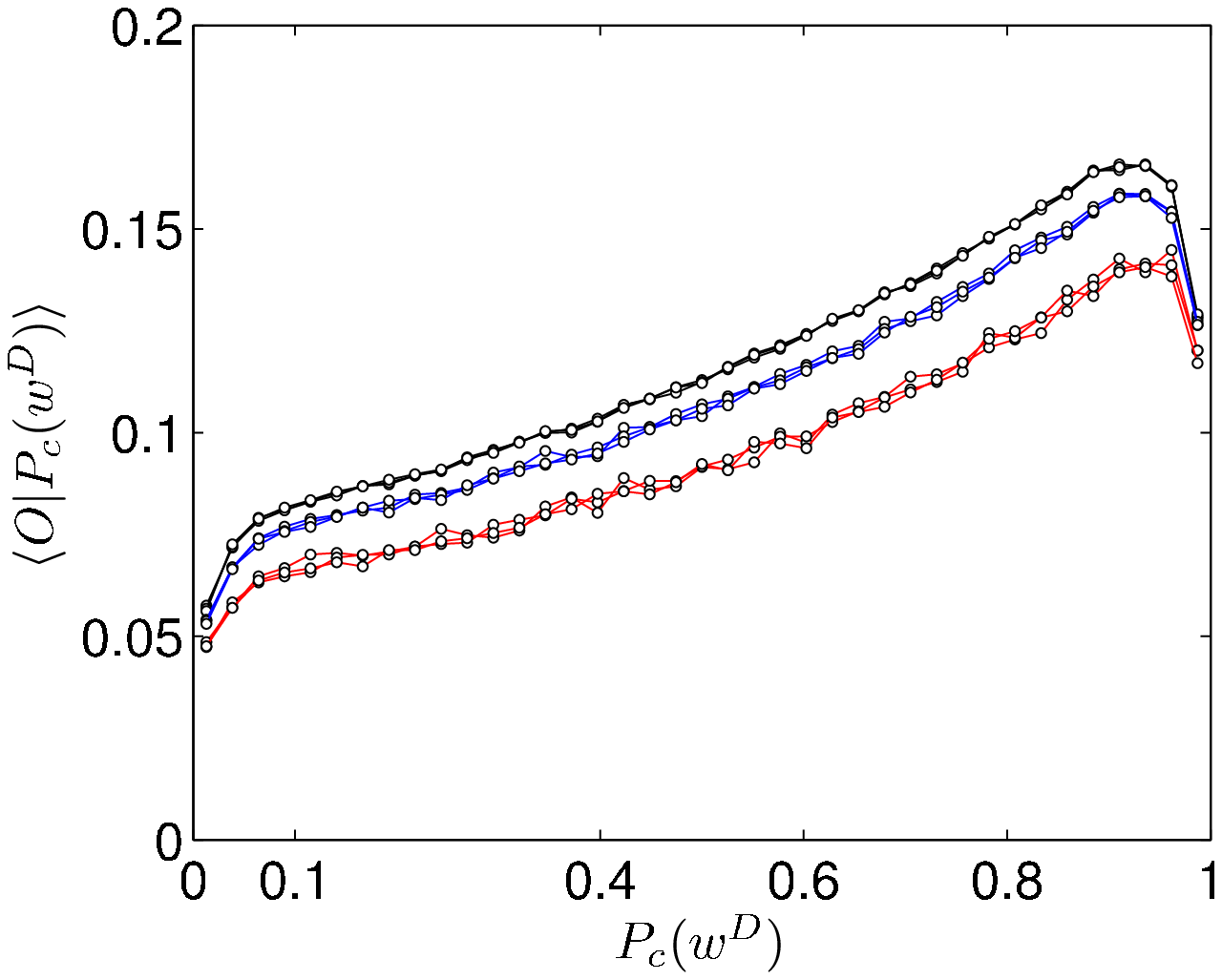}
\caption{Average link overlap as a function of link weight $\langle O|w^D,p \rangle$ (left) and cumulative link weight $\langle O|P_{c}(w^D),p \rangle$ (right) for altogether nine network samples for the LCC of the mutual network. Three samples were drawn for each value of $p$, corresponding to the probability of a node in the initial network to be included in the sample. We used the values of $p=0.8$ (top 3 curves), $p=0.6$ (middle 3 curves), and $p=0.4$ (bottom 3 curves), and the corresponding sample sizes were $N_{p=0.8} \approx 2.6 \times 10^6$, $N_{p=0.6} \approx 1.4 \times 10^6$, and $N_{p=0.4} \approx 0.4 \times 10^6$.}
\label{fig:resample}
\end{center}
\end{figure}

To estimate the effect of $p$ on $\langle O | w^D\rangle$, we follow an approach motivated by the Bootstrap-technique \cite{efron:1994}. We generate a resample of the LCC of our network by including each node in the resample with probability $p$ and by varying it obtain different sample sizes. In the limit of setting $p=1$ we recover the original network. The results are shown in Fig.~\ref{fig:resample}. Although lower values of  $p$ result in slightly lower values of $\langle O | w^D\rangle$, its qualitative behavior is fairly insensitive to it. The cumulative plot shows how decreasing $p$ does, in fact, cause the curve to become slightly flatter. This suggests that if the original network covered a larger fraction of the market or, alternatively, if data from several phone operators was aggregated, the value of $\langle O|P_{c}(w^D) \rangle$ would somewhat increase in absolute terms but, most importantly, its increasing trend with respect to $w^D$ would become possibly even more pronounced. In short, the reported relationship between weight $w$ and overlap $O$ is not an artifact caused by having a sample from the underlying mobile phone call network.

A well-known hypothesis from sociology, the weak ties hypothesis of Granovetter, states that the proportional overlap of two individual's friendship networks varies directly with the strength of their tie to one another \cite{granovetter:1973}. According to this hypothesis, the strength of a tie is a ``combination of the amount of time, the emotional intensity, the intimacy (mutual confiding), and the reciprocal services which characterize the tie'' \cite{granovetter:1973}. The present network is suitable for testing the weak tie hypothesis empirically at a societal level for two reasons. First, the weights are phone call durations and thus implicate the time commitment to the relationship, one of the variables suggested to be indicative of the strength of an interpersonal tie. Second, the size of the network guarantees sufficient averaging and, therefore, produces reliable statistics. In addition, using the non-mutual network entails at least some degree of reciprocity (at least one call has been returned) and, importantly, commitment of phone time in this case also implies monetary costs to the caller. The average overlap increases for about 95\% of link weights, as shown in Fig.~\ref{fig:overlap}, and the behavior of the remaining 5\% can be accounted for (see Supplementary Material in \cite{onnela:2007}). Importantly, this increasing trend is practically unaffected whether number of calls $w^N$ or aggregate call duration $w^D$ are used as weights. Put together with the issue of sampling discussed above, these results provide a societal level verification of the weak ties hypothesis \cite{onnela:2007}.  

\begin{figure}
\begin{center}
\includegraphics[width=0.45\linewidth]{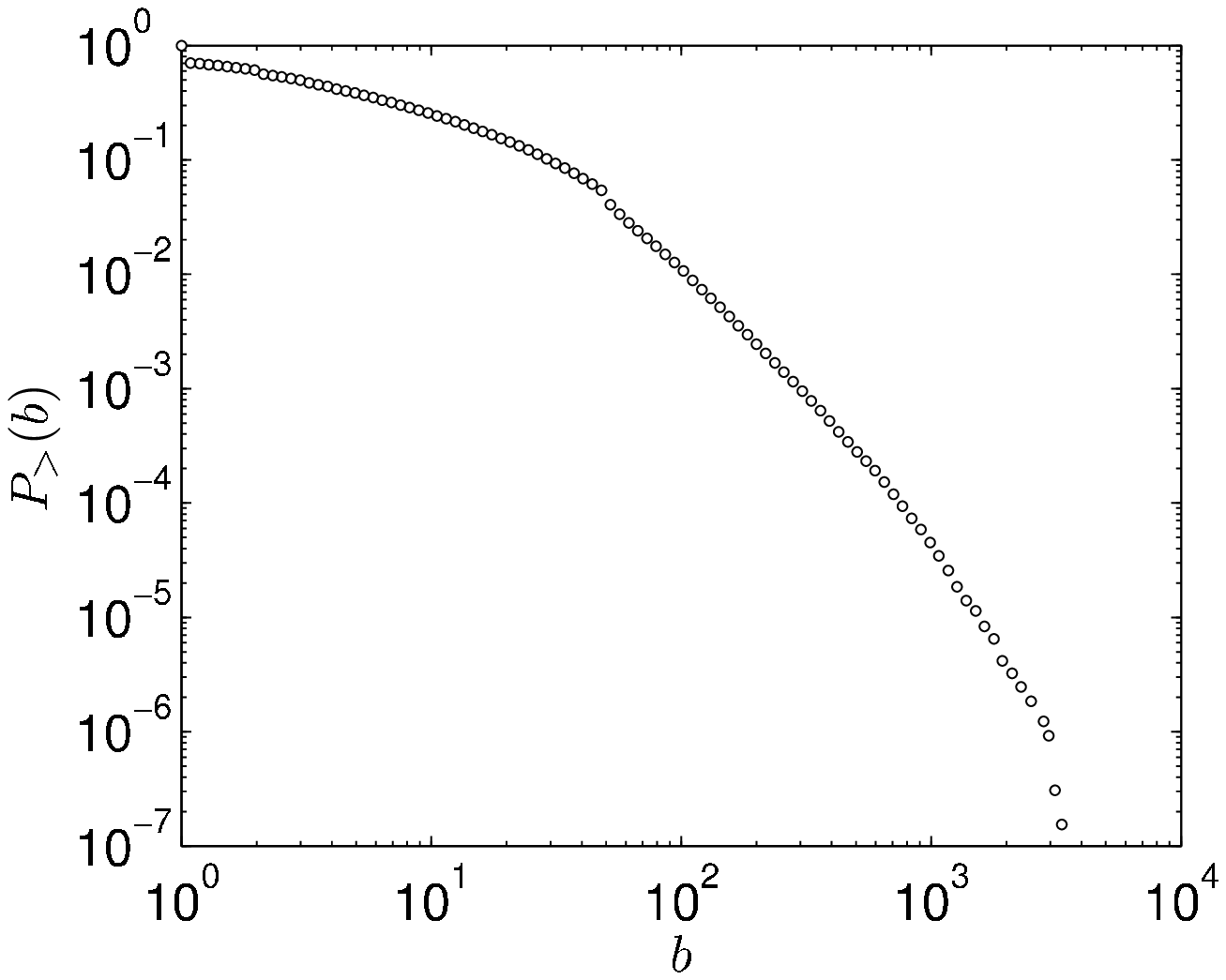}
\includegraphics[width=0.45\linewidth]{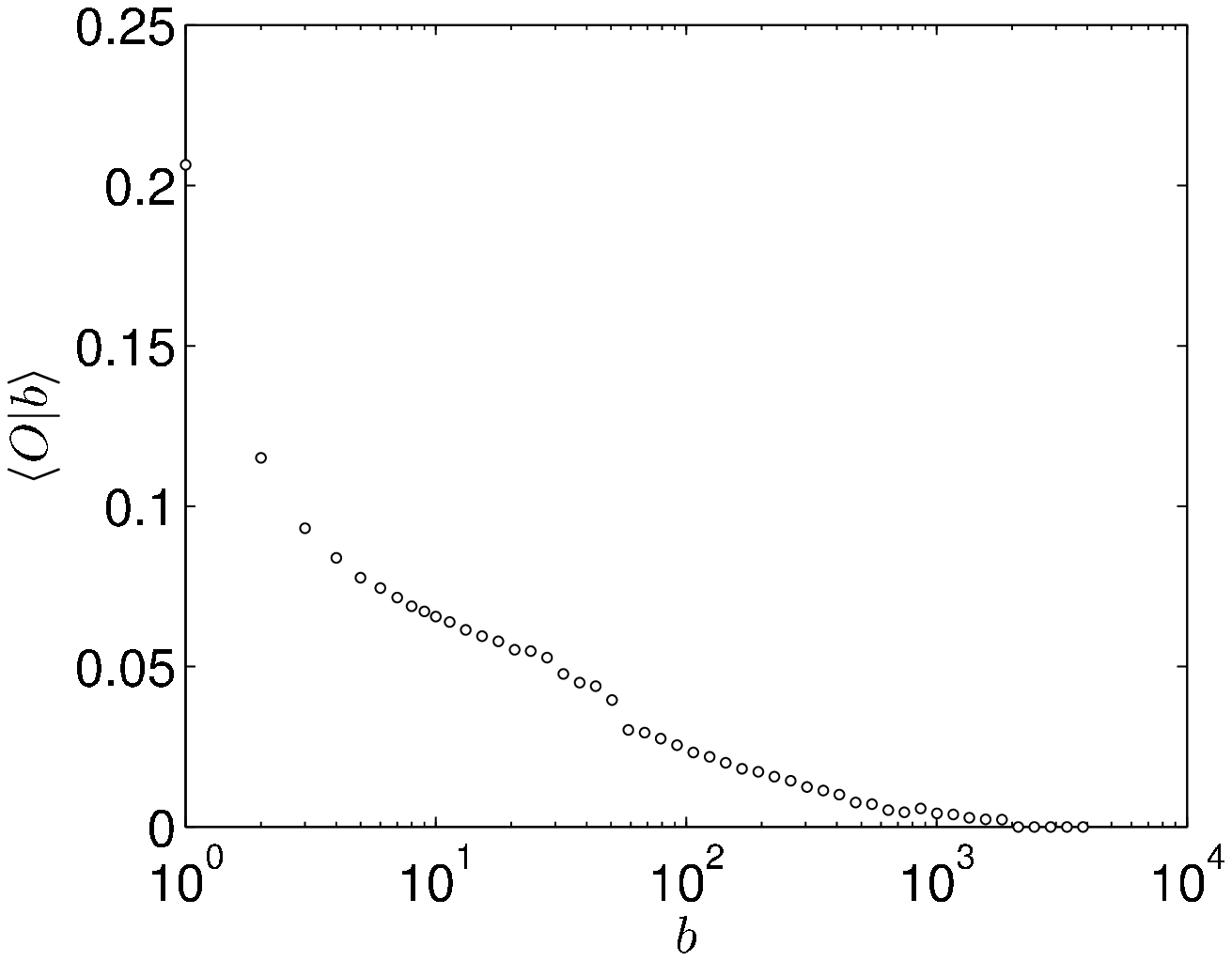}
\caption{Ccumulative distribution of link betweenness centrality $P_{>}(b)$ (left) in the LCC of the mutual network and the average link overlap as a function of link betweenness centrality $\langle O|b \rangle$ (right).  Here $P_{>}(b)$ has been computed using a sample of $N_s = 10^5$ starting nodes from which the shortest paths to every other $N-1$ nodes were found in order to calculate the betweenness centrality of links.}
\label{fig:betcent}
\end{center}
\end{figure}

The results on overlap can be related to the concept of link betweenness centrality, defined for a link $e=(i,j)$ as $b_{ij}  = \sum_{v \in V_s} \sum_{w \in V / \{v\}} \sigma_{vw}(e) / \sigma_{vw}$, where $\sigma_{vw}(e)$ is the number of shortest paths between $v$ and $w$ that contain $e$, and $\sigma_{vw}$ is the total number of shortest paths between $v$ and $w$ \cite{holme:2002d}. In practice, we use the algorithm introduced in \cite{newman:2001b} to compute $b_{ij}$ but, due to the heavy computational requirements of the algorithm, instead of using all the nodes of the set $V$ making up the network, we use a subset of $N_s = 10^5$ nodes in the sample $V_s$ as starting points. We then use the algorithm to find the shortest paths from these $N_s$ nodes to all other remaining $N-1$ nodes, every time keeping track of which links are used in constructing the shortest paths. Note that using this many source nodes results of the order of $10^{11}$ shortest paths to be computed in the network, more than a sufficient number, as was confirmed by using a smaller value for $N_s$. The cumulative distribution of link betweenness centrality is shown in Fig.~\ref{fig:betcent}. The figure also shows the behavior of average link overlap as a function of link betweenness centrality $\langle O|b \rangle$. This is in full agreement with the above picture of the role of weak and strong links: Weak links have low overlap but high betweenness centrality, reflecting their importance in holding the system together, while strong links have high overlap but low betweenness centrality and, as such, unlike the weak links, are not irreplaceable.

\section{Percolation studies}
\label{sec:perc}

We now turn to an examination of the implications of link removal on the global properties of networks, which has many precedents in the complex network literature \cite{cohen:2000,holme:2002d,broder:2000,albert:2000,callaway:2000,cohen:2001,watts:2002,holme:2002c,onnela:2004}. However, instead of removing links randomly, we remove them based on either their weight $w_{ij}$, overlap $O_{ij}$, or betweenness centrality $b_{ij}$ values. Removal can be carried out in one of two directions, i.e., either starting from links with low $w_{ij}$, $O_{ij}$, or $b_{ij}$ values and proceeding towards higher ones or, alternatively, starting from links with high $w_{ij}$, $O_{ij}$, or $b_{ij}$ and proceeding towards those with lower corresponding values. This thresholding process is governed by the control parameter $f$, the ratio of removed links, which allows us to interpolate between the initial connected network ($f=0$) and a set of isolated nodes ($f=1$). We study the response of the network to removal of $w_{ij}$, $O_{ij}$, and $b_{ij}$ links by monitoring four quantities as a function of the control parameter, which are (1) order parameter $R_{\textrm{LCC}}$, the fraction of nodes in the LCC, (2) 'susceptibility' $\tilde{S} = \sum_{s} s^2 n_s$, where $n_s$ is the number of clusters of size $s$, and (3) average shortest path length $\langle \ell \rangle$. In addition, we also study the effect of link removal on the (4) average clustering coefficient $\langle C \rangle$. Differences in the behavior of these quantities reflect the global role different links have in the network.

The order parameter $R_{\textrm{LCC}}$ is defined as the fraction of nodes in the LCC, i.e., the fraction of nodes that can all reach each other through connected paths. We find that removing links from low $w_{ij}$ to high $w_{ij}$ (red curve), from low $O_{ij}$ to high $O_{ij}$ (red curve), or from high $b_{ij}$ to low $b_{ij}$ (black curve) leads to a sudden disintegration of the network at $f^w=0.8$, $f^O=0.6$, and $f^b=0.6$, respectively. In contrast, removing first the high weight, high overlap, or low betweenness centrality links will shrink the network, but will not precipitously break it apart. This suggests that weak and strong links, low and high overlap links, and low and high betweenness centrality links have all different global structural roles in the network. In particular, it appears that removing low overlap links produces a qualitatively similar response to removing high betweenness centrality links.

\begin{figure}
\begin{center}
\includegraphics[width=1.0\linewidth]{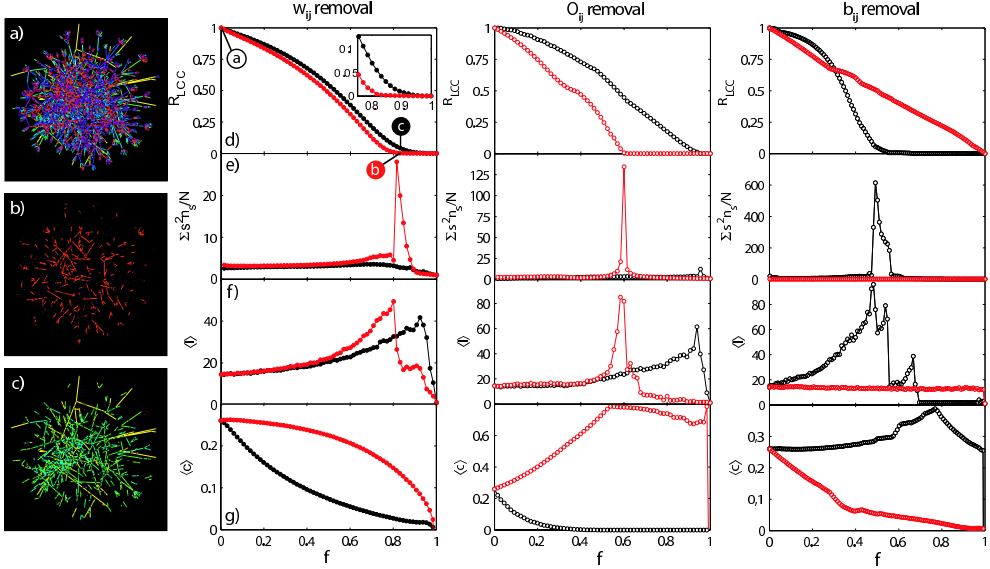}
\caption{Percolation analysis. Panel (a) shows a  small network sample with all links intact, (b) the same sample with 80\% of the low $w_{ij}$ links removed ($f=0.80$, red curve), and (c) the sample with 80\% of high $w_{ij}$ links removed ($f=0.80$, black curve). Rows (d, e, f, g): Removal of high or low weight $w_{ij}$ links (left column), overlap $O_{ij}$ links (middle column), or betweenness centrality $b_{ij}$ links (right column). The links are removed one at a time based on their ranking, such that the black curves correspond to starting removal from high $w_{ij}$, $O_{ij}$, and $b_{ij}$ links, whereas the red curves represent the opposite, starting removal from low $w_{ij}$, $O_{ij}$, and $b_{ij}$ links. The fraction of removed links is denoted by $f$. Row (d): The order parameter $R_{\textrm{LCC}}$, the fraction of nodes of nodes present in the LCC of the network for the given value of $f$ to that present in the LCC for $f=0$. Row (e): $\tilde{S} = \sum_s s^2 n_s / N$, corresponding to the average component size in the network with the LCC excluded from the summation. Row (f): Average shortest path length $\langle \ell \rangle$ in the LCC of the system for the given value of $f$, which is also expected to diverge as $f \to f_c$. Row (g): Average clustering coefficient $\langle C \rangle$ in the network.}
\label{fig:percolation}
\end{center}
\end{figure}

The second row shows the behavior of $\tilde{S} = \sum_s s^2 n_s / N$, which is analogous to magnetic susceptibility in thermal phase transitions, corresponding to the average component size in the network with the LCC excluded from the summation. According to percolation theory, if the network collapses via a phase transition at $f_c$, then $\tilde{S}$ diverges as $f \to f_c$ for an infinite system. A finite signature of such divergence is clearly visible in these plots upon removing low $w_{ij}$, low $O_{ij}$, or high $b_{ij}$ links, suggesting that the network disintegrates at this point following a phase transition. Since the role of weak and strong ties is different at the local level and has important consequences from the sociological perspective \cite{granovetter:1973}, understanding their different global role is central, which is indeed a very pertinent question from the perspective of social network theory (see Section \ref{sec:intro}). We have studied the global role of weak and strong links using finite size scaling (FSS) as reported in \cite{onnela:2007}. Although different FSS methods yielded slightly different results, removal of weak links (red curve) lead  to a genuine phase transition at around $f_c^w(\infty)=0.80$, but there appears to be no phase transition when strong links are removed first (black curve). This result indicates  that weak and strong links have qualitatively different global roles in social networks.

While the size of the largest component tells us about overall connectivity of the network, it does not convey information about its topology, only that the $N_{\textrm{\tiny{LCC}}}(f=0) R_{\textrm{\tiny{LCC}}}(f)$ nodes are connected through one or more paths. One way to characterize the topology of the network is to study the average shortest path length, denoted by $\langle \ell \rangle$, which is the average number of links on the shortest path connecting any two vertices within the LCC. Note that as links are removed, the network becomes fragmented in components, of which we focus only on the largest one, i.e., the LCC for the given value of the control parameter $f$. Path lengths are also important  from the perspective of network function and efficiency. The existence of a path between nodes is a necessary but not sufficient condition for there to be a flow of information between them. This is especially true if the transmission through links is leaky, i.e. it is possible for information to get lost along the way. Focusing on the role of weak and strong ties, we find that removal of weak ties increases path lengths more than removal of strong ties does, although the effect is stronger upon removing low $O_{ij}$ or high $b_{ij}$ links.

Path lengths are also related to the conjecture obtained from the weak ties hypothesis, according to which communities are locally connected by single weak ties, and removing these weak ties should therefore increase average path lengths making it more difficult to reach people \cite{granovetter:1973}. Our result provides an empirical verification of the weak ties conjecture. It can also be related to a study dealing with search in social networks, according to which successful searches are conducted primarily through intermediate to weak strength ties without requiring highly connected hubs to succeed \cite{dodds:2003}. The present results suggest that the success of weak ties for search might lie in their function as community connectors, enabling one to reach outside of one's own community and thus expanding the set of individuals who may be reached through the network.

The average clustering coefficient $\langle C \rangle$ measures the local cliquishness of the network. Unlike the average shortest path length $\langle \ell \rangle$, which is computed only for the LCC for the given value of $f$, the average clustering coefficient is computed over all nodes in the network for which degree $k>1$. Removing strong links (Fig.~\ref{fig:percolation}, row (g), black curve) leads to a convex clustering curve with an overall lower $\langle C \rangle$ than when weak links are removed. This happens because the strong links are mostly located in tightly connected communities where triangles are abundant. Consequently, removing them decreases the number of triangles and lowers clustering. Removing weak links (red curve) produces a concave clustering curve which first decreases very slowly. This is because the weak links are mostly located between communities, acting as local bridges and, therefore, rarely participate in triangles. Consequently, removing them has little effect on clustering. However, the difference in behavior for overlap thresholding is even more drastic. On removing high $O_{ij}$ links, the communities become shattered very quickly, so that at $f \approx 0.40$ average clustering coefficient is close to zero. The opposite happens on removing low $O_{ij}$ links. The average clustering increases up-to $f \approx 0.54$, compatibly with the fact that 53.5\% of links in the GC have $O_{ij}=0$, and reaches a value almost as high as $\langle C \rangle \approx 0.80$. This results demonstrates quantitatively that the network is highly clustered and these clusters, or communities, can be filtered out reasonably well by removing low $O_{ij}$ links. Again, removal of high overlap links is again qualitatively similar to removing low betweenness centrality links and vice versa.

Since some community detection algorithms rely on the concept of betweenness centrality to detect communities \cite{girvan:2002}, our results suggest that it may be possible to use the concept of overlap to detect communities at least in social networks. Bearing in mind that $O_{ij}$ is a local characteristic and can be computed in $O(N)$, whereas $b_{ij}$ is a global characteristic and takes $O(N^2 \ln{N})$ to compute, algorithms relying on $b_{ij}$ could use $1/O_{ij}$ as a local proxy for $b_{ij}$, potentially leading to significant gains in computing performance. We note that a modified version of the edge-clustering coefficient of Eq.~\ref{eq:linkclust} has also been used to replace edge betweennes centrality in a popular method for finding communities \cite{radicchi:2004a}. One could alternatively use $O_{ij}$ without any modifications and, due to its desirable properties covered in Section \ref{sec:link}, it may be better suited for that purpose in identifying communities in social networks.

\section{Discussion}
\label{sec:disc}

Modern technologies enable the study of social networks of unprecedented size. A number of such investigations have appeared recently ranging from exploring email communication networks \cite{ebel:2002,eckmann:2004,dodds:2003,wilkinson:2003} to identifying groups and strategies in an electronic  marketplace \cite{yang:2006,yang:2006a,reichardt:2005}. In this paper we constructed a network from mobile phone call records and used both aggregated call durations and the cumulative number of calls as a measure of the strength of a social tie. Since the network is derived exclusively from one-to-one communication, it can be used as a proxy for the underlying human communication network at the societal level which, to our knowledge, is the largest weighted social network studied as far.
 
In prototypical sociological studies the number of investigated individuals is limited to the order of hundred \cite{giddens:2006}, although exceptionally, like in the case of the Add Health database \cite{addhealth} as employed, for example, in \cite{gonzalez:2007}, tens of thousand of individuals may be reached using questionnaires. This method enables covering a broad spectrum of interpersonal relations, although the subjectivity and quantification of interaction strengths are major problems. In this paper we have followed a complementary approach by basing the network on a specific type of social interaction, a phone call, allowing an objective measure of interactions for millions of people. We believe that studies like this one can provide valuable lessons about the large-scale structure of societies emerging from microscopic social interactions.

One of our focal points was to explore the relationship between local network topology and the associated weights. This is particularly important from the point of view of  sociology, where such a relation has been hypothesized a long time ago. In order to test the weak ties hypothesis, we used the concept of link overlap to measure the coupling between link weight and the overlap of the neighborhood in the vicinity of the tie. We demonstrated that for 95\% of the links the overlap and tie strength are correlated, verifying the hypothesis at a societal level. Moreover, we found the link overlap to be negative correlated with its betweenness centrality, suggesting that the former can be used as a local proxy for the latter, computationally heavy, global quantity.

We explored further the role of weights in the network using the concepts of intensity, coherence, and weighted clustering coefficient. We found correlations between local network structure at the level of cliques, or communities, and interactions strengths within them. The weighted clustering coefficient provides an appropriate tool for probing the strength of clustering due to weights, and may be used to differentiate between weighted networks that have fundamentally different coupling between network topology and interaction strengths. We found that the network is assortative in terms of topology as expected but, rather surprisingly, is not weight-assortative for a large majority of nodes. Further, the coupling between local network structure and interaction strengths carries over to the global level. We quantified this by studying the differences in percolation behavior depending on the properties of the removed links. Following this approach we also verified the so-called weak ties conjecture, a global manifestation of the weak ties hypothesis.

The obtained results can be used as a basis for devising weighted models of social networks. In particular, the relation between topological and statistical properties should be incorporated in such models. This enables studying collective social phenomena, such as spreading of information and opinion formation, at a level of realism and scale not possible in the past. The lessons learnt from this endeavor are not limited to understanding human societies, but may find application in other domains as well. Finally, we believe that our systematic approach can be adopted to study other weighted networks, and the present results can bee seen as a reference against which other networks may be compared. 

\vspace{1cm}

\bibliographystyle{unsrt}

\vspace{1cm}

\textbf{Acknowledgments}: JPO is grateful to the ComMIT Graduate School, and Finnish Academy of Science and Letters, V\"ais\"al\"a Foundation, for a travel grant to visit A.-L. Barab\'{a}si at Harvard University. This research was partially supported by the Academy of Finland, Centres of Excellenc programmes, project no. 44897 and 213470 and OTKA K60456. GS and ALB were supported by the NSF-ITR, NSF-DDAAS projects and by the McDonell Foundation.

\end{document}